\documentclass[aps,twocolumn,showpacs,superscriptaddress,groupedaddress]{revtex4}  
\usepackage{graphicx}  
\usepackage{dcolumn}   
\usepackage{bm}        
\usepackage{amssymb}   
\usepackage{amsmath}   
\usepackage{braket}
\usepackage{footnote}
\usepackage{subfigure}
\usepackage{color}
\usepackage{here}
\usepackage{hyperref}
\usepackage{multirow}
\usepackage{enumitem}

\usepackage[T1]{fontenc}

\newcommand{\Lagr}{\mathcal{L}}

\usepackage{color}
\usepackage[normalem]{ulem}

\usepackage[T1]{fontenc}

\begin{document}
\title{\boldmath Leptogenesis from oscillations and dark matter}



\author{A. Caputo}
\affiliation{IFIC (CSIC-UVEG), Edificio Institutos Investigaci\'on, 
Apt.\ 22085, E-46071 Valencia, Spain}
\author{P. Hern\'andez}
\affiliation{IFIC (CSIC-UVEG), Edificio Institutos Investigaci\'on, 
Apt.\ 22085, E-46071 Valencia, Spain}
\author{N. Rius}
\affiliation{IFIC (CSIC-UVEG), Edificio Institutos Investigaci\'on, 
Apt.\ 22085, E-46071 Valencia, Spain}


\begin{abstract}
An extension of the Standard Model with 
	Majorana singlet fermions  in the 1-100~GeV range can explain the light neutrino masses and give rise to a baryon asymmetry  at freeze-in of the heavy states, via their CP-violating oscillations. 
	In this paper we consider extending this scenario to also explain dark matter.  We find that a very weakly coupled $B-L$ gauge boson, an invisible QCD axion model, and the singlet majoron model can simultaneously account for dark matter and the baryon asymmetry. 
	\end{abstract}
\maketitle

\section{Introduction}
\label{sec:intro}

The Standard Model (SM) of particle physics needs to be extended to explain neutrino masses, the missing  gravitating matter (DM)  and the observed matter-antimatter asymmetry in the universe. 

Some of the most minimal extensions of the SM include new fermions, namely two or three sterile Majorana neutrinos (singlets under  the full gauge group), which can account for the tiny 
neutrino masses, through the seesaw mechanism \cite{Minkowski:1977sc,GellMann:1980vs,Yanagida:1979as,Mohapatra:1979ia}, and explain the observed matter-antimatter asymmetry  through leptogenesis \cite{Fukugita:1986hr}. The simplest version of leptogenesis establishes the source of the matter-antimatter asymmetry in the CP violating the out-of-equilibrium decay of the heavy neutrinos. This scenario requires however relatively large Majorana masses $> 10^8$ GeV \cite{Davidson:2002qv} (or $\sim 10^6$ GeV with flavour effects \cite{Abada:2006ea} included), which makes these models difficult to test experimentally.  For Majorana neutrinos in the 1-100 GeV range, it has been shown by Akhmedov, Rubakov
and Smirnov (ARS) \cite{Akhmedov:1998qx} and refined by Asaka and Shaposhnikov (AS) \cite{Asaka:2005pn} that a different mechanism of leptogenesis is at work. In this case the asymmetries are produced at freeze-in of the sterile states via their CP-violating oscillations. The original ARS proposal did not include flavour effects and needs at least three Majorana species, while AS have shown that flavour effects can make it work with just two species. In the rest of the paper we will refer indistinctively to both scenarios as baryogenesis  from oscillations (BO).  In both cases, the lepton asymmetry is reprocessed into a baryonic one by electroweak sphalerons \cite{Kuzmin:1985mm}. The extra heavy neutrinos in this case could be produced and searched for in beam dump experiments and colliders (see \cite{Ferrari:2000sp,Graesser:2007pc,delAguila:2008cj,BhupalDev:2012zg,Helo:2013esa,Blondel:2014bra,Abada:2014cca,Cui:2014twa,Antusch:2015mia,Gago:2015vma,Antusch:2016vyf,Caputo:2016ojx,Caputo:2017pit} for an incomplete list of works), possibly giving rise to spectacular signals such as displaced vertices \cite{Helo:2013esa,Blondel:2014bra,Cui:2014twa,Gago:2015vma,Antusch:2016vyf}. 
Since only two sterile neutrinos are needed to generate the baryon asymmetry \cite{Asaka:2005pn,Shaposhnikov:2008pf,Canetti:2012zc,Canetti:2012kh,Asaka:2011wq,Shuve:2014zua,Abada:2015rta,Hernandez:2015wna,Hernandez:2016kel,Drewes:2016gmt,Drewes:2016jae,Hambye:2016sby,Ghiglieri:2017gjz,Asaka:2017rdj,Hambye:2017elz,Abada:2017ieq,Ghiglieri:2017csp}, the  lightest sterile neutrino in the keV range can be very weakly coupled and play the role of DM \cite{Dodelson:1993je}. This is the famous $\nu$MSM \cite{Asaka:2005pn}. However the stringent X-ray bounds imply that this scenario can only work in the presence of a leptonic asymmetry \cite{Shi:1998km} significantly larger than the baryonic one, which is quite difficult to achieve. A recent update of astrophysical bounds on this scenario can be found in \cite{Perez:2016tcq,Baur:2017stq}

In this paper our main goal is to consider scenarios compatible with Majorana masses in the 1-100 GeV range and study the conditions under which the models can explain DM without spoiling ARS leptogenesis \footnote{  Some very recent work  along these lines in the scotogenic model  can be found in \cite{Baumholzer:2018sfb}}. 
In particular, we will focus on models that are minimal extensions of the type I seesaw model with three singlet neutrinos. We will first consider an extension involving a gauged $B-L$ model \cite{Mohapatra:1980qe}, which includes an extra gauge boson and can explain DM in the form of a non-thermal keV neutrino. We will then consider an extension which includes a CP axion \cite{Langacker:1986rj} that can solve the strong CP problem and  explain DM in the form of cold axions. Finally we consider the majoron singlet model \cite{Chikashige:1980ui,Schechter:1981cv} which can also explain DM under certain conditions both   in the form of a heavy majorana neutrino or a majoron. 

The plan of the paper is as follows. We start by briefly reviewing the ARS mechanism and the essential ingredients and conditions that need to be met when the sterile neutrinos have new interactions. In section \ref{B-L} we discuss the gauged $B-L$ model,  in section \ref{axion}, we study the invisible axion model with sterile neutrinos and in section \ref{majoron} we reconsider the singlet majoron model. In section \ref{conclusion} we conclude.

\section{Leptogenesis from oscillations}
For a recent extensive review of the ARS mechanism see \cite{Drewes:2017zyw}.  The model is just the type I seesaw model with three 
neutrino singlets, $N_i$, $i =1- 3$, which interact with the SM only through their Yukawa couplings. The Lagrangian in the Majorana mass basis is
\begin{equation}
{\mathcal L}= {\mathcal L}_{\rm SM} +i  \overline{N}_{ i}\gamma^{\mu}\partial_{\mu} N_{ i}\\
- \left(Y_{\alpha i} \overline{L}_\alpha N_{i}\Phi + \frac{m_{N_{i}}}{2} \overline{N}^c_{ i} N_{ i} +h.c.\right).
\label{eq:seesaw}
\end{equation}
In the  early Universe before the electroweak (EW) phase transition, the singlet neutrinos are produced through their Yukawa couplings in flavour states, which are linear combinations of the mass eigenstates. Singlet neutrinos then oscillate, and since CP is not conserved, lepton number $L$ gets unevenly distributed between different flavours. At high enough temperatures $T \gg m_{N_{i}}$,  total lepton number vanishes, in spite of which a surplus of baryons over antibaryons can be produced, because the flavoured lepton asymmetries are stored in the different species and transferred at different rates to the baryons. 
As long as full equilibration of the sterile states is not reached before the EW phase transition ($T_{\rm EW} \sim 140$GeV) , when sphaleron processes freeze-out, a net baryon asymmetry survives.  It is {\it essential} that at least one of the sterile neutrinos does not  equilibrate by $t_{\rm EW}$.   The rate of interactions of these neutrinos at temperatures much higher than their mass can be estimated to be
\begin{eqnarray}
\Gamma_\alpha \propto \kappa y_{\alpha}^2 T, 
\end{eqnarray}
where $ y_\alpha$ are the eigenvalues of the neutrino Yukawa matrix, $T$ is the temperature and $\kappa =$ few $10^{-3}$ \cite{Besak:2012qm,Garbrecht:2013urw,Ghisoiu:2014ena}. 
The Hubble expansion rate in the radiation dominated era is
\begin{eqnarray}
H(T) =  \sqrt{4 \pi^3 G_N g_* \over 45} T^2 \equiv {T^2 \over M_{P}^*}.
\end{eqnarray}
where $g_*$ is the number of relativistic degrees of freedom ($g_*\sim 100$ above the EW phase transition). 
The requirement that no equilibration is reached before $t_{\rm EW}$ is:
\begin{eqnarray}
\Gamma_\alpha(T_{\rm EW}) \leq H(T_{\rm EW}),
\end{eqnarray}
which implies yukawa couplings of order
%
\begin{equation}
y_\alpha \lesssim 10^{-7},
\label{eq:y}
\end{equation}
i.e. not much smaller than the electron yukawa. These yukawa couplings 
are compatible with the light neutrino masses for Majorana masses in the 1 GeV-100 GeV range.

Any model that extends the one described above with new fields/interactions should be such that the new interactions do not increase the equilibration rate of the sterile neutrinos for the out-of-equilibrium requirement in the ARS mechanism to be met. We will now consider the implications of this requirement on various extensions of the minimal seesaw model of eq.~(\ref{eq:seesaw}) that are well motivated by trying to explain also the dark matter and in one case also the strong CP problem.

\section{B-L gauge symmetry}
\label{B-L}

The SM is  invariant under an accidental global $U(1)_{B-L}$ symmetry, that couples to baryon minus lepton number. If one promotes this symmetry to a local one \cite{Mohapatra:1980qe},  the model needs to be extended with three additional right handed neutrinos to avoid anomalies, which interestingly makes the type I seesaw model the minimal particle content compatible with this gauge symmetry. In this case, we have interactions between SM lepton and quark fields with the new gauge boson, $V_\mu$, as well as an additional term involving sterile neutrinos
\begin{equation}
{\mathcal L} \supset g_{B-L} \left( \sum_f Q^f_{B-L}  V_\mu \overline{f} \gamma_\mu  f - \sum_a V_{\mu}\overline{N}_a\gamma^{\mu}N_a \right),
\end{equation}
where $Q_{B-L}^f = 1/3, -1$ for quarks and leptons respectively. We also assume the presence of a scalar field $\phi$, with charge $B-L$ charge 2:
\begin{eqnarray}
\Lagr \supset (D_\mu \phi)^\dagger D_\mu \phi -V(\phi) - {h_N\over 2}\overline{N}^c N \phi + h.c. ,
\end{eqnarray}
that gets an expectation value $\langle \phi \rangle$, breaking $B-L$ spontaneously \footnote{The Stuckelberg mechanism cannot be used here, because we need heavy neutrinos.}, and giving a mass to both the gauge boson and the sterile neutrinos:
\begin{equation}
m_V = 2 \sqrt{2} g_{B-L}\langle\phi\rangle, \;\;\; m_{N_i} = h_{N_i} \langle\phi\rangle.
\label{eq:mv}
\end{equation}
A massive higgs from the $B-L$ breaking, $\sigma$, remains in the spectrum with a mass that we can assume to be $M_\sigma \sim \langle \phi\rangle$.

\begin{figure}
	\centering
	\includegraphics[width=8.cm,keepaspectratio]{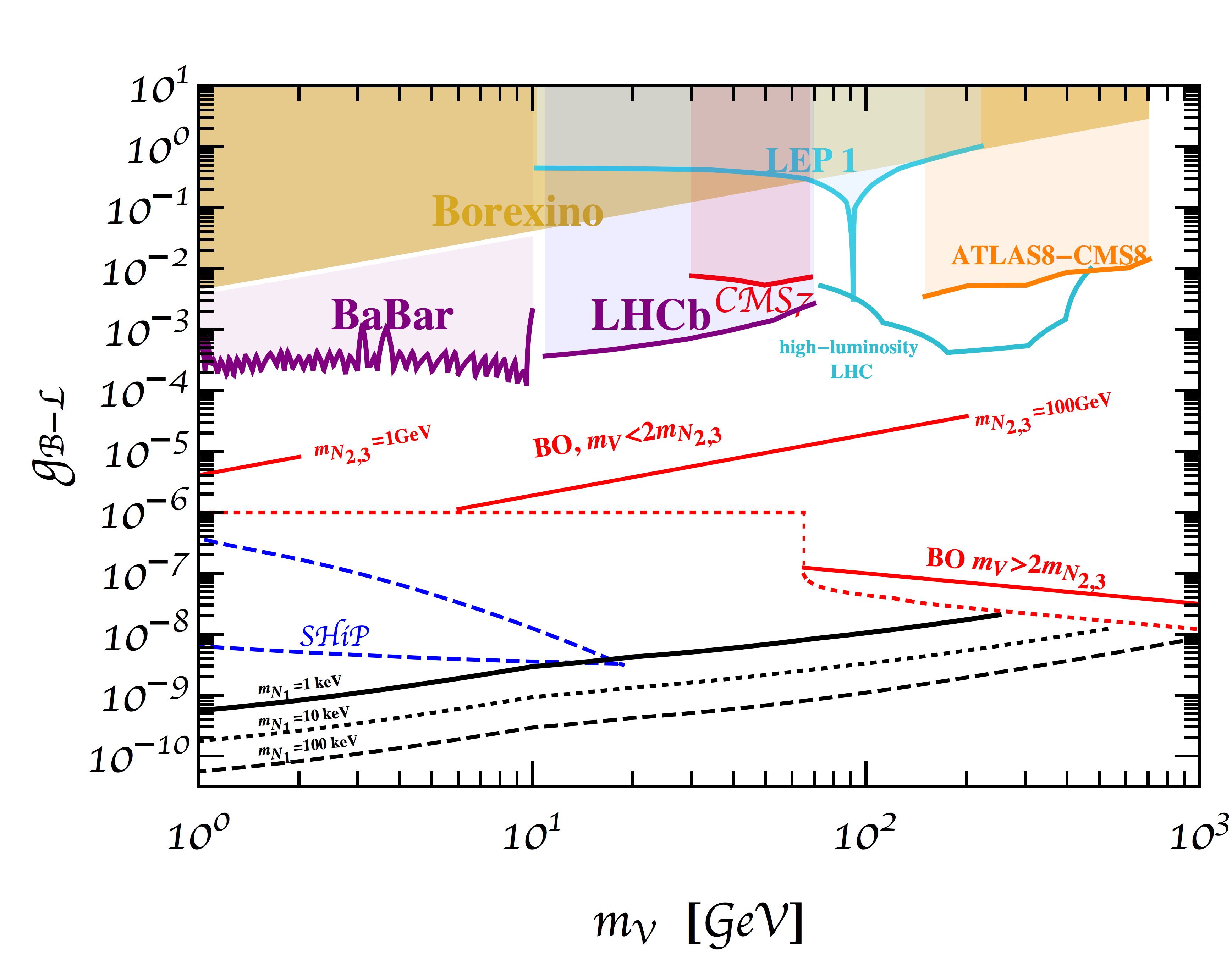}
	\caption{Summary of present (shaded regions) and future (unshaded) constraints on $g_{B-L}$ and $m_V$ for the $B-L$ model, adapted from ref.~\cite{Batell:2016zod}. The dashed region labeled SHIP is the sensitivity of bremsstrahlung searches in SHIP  \cite{Gorbunov:2014wqa,Kaneta:2016vkq}. The solid, dotted and dashed black lines correspond to the correct DM relic abundance in the form of sterile neutrinos of mass 1, 10 and 100 keV respectively. The dotted red line indicates the lower limit on the gauge coupling for which $V$ is in thermal equilibrium. The solid red lines correspond to the upper bounds for successful BO leptogenesis for $m_{N_{2,3}} =1$ and 100 GeV.  }
	\label{fig:bound}
\end{figure}

Existing constraints on this model come from direct searches for $V$  in elastic neutrino-electron scattering,  $V$ gauge boson production at colliders, Drell-Yan processes and new flavour changing  meson decays \cite{Bellini:2011rx,Chatrchyan:2013tia,Lees:2014xha,Lees:2017lec,Khachatryan:2014fba,Aad:2014cka,ALEPH:2004aa,Appelquist:2002mw}. The status of these searches is summarized in Fig.~\ref{fig:bound},  adapted from \cite{Batell:2016zod} (see also \cite{Klasen:2016qux,Ilten:2018crw,Escudero:2018fwn}). For masses,  1 GeV $\leq m_V\leq10$ GeV, $g_{B-L}$ is bounded to be  smaller than $\sim 10^{-4}$, while the limit is weaker for larger masses. 
The improved prospects to search for right-handed neutrinos exploiting the $U(1)_{B-L}$ interaction have been recently studied in \cite{Batell:2016zod}, where the authors consider the displaced decay of the $N$ at the LHC and the proposed SHIP beam dump experiment\cite{Alekhin:2015byh}. 

\begin{figure}
	\centering
	\includegraphics[width=8cm]{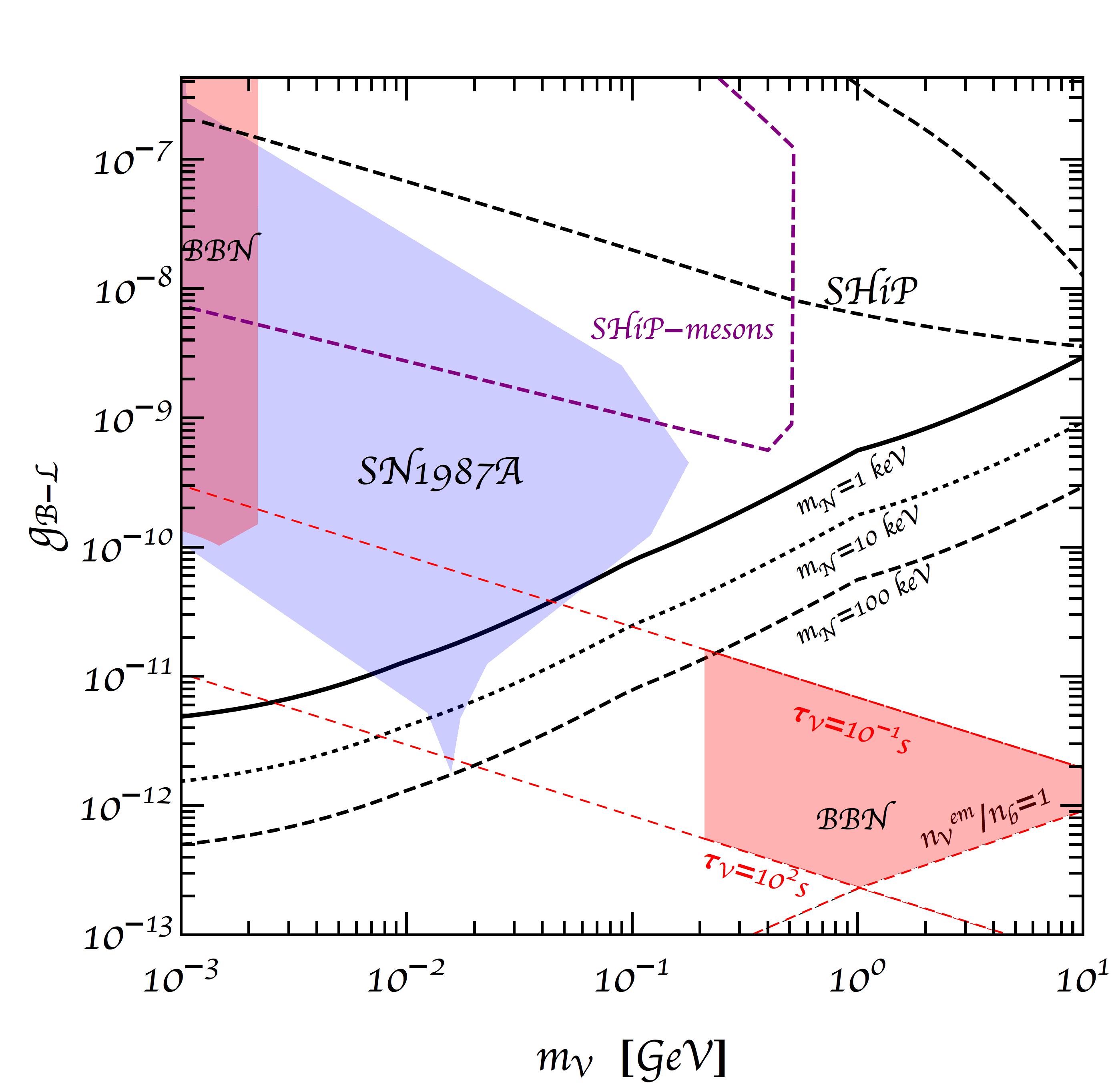}
	\caption{Shaded regions are presently excluded by supernova observations \cite{Chang:2016ntp},  BBN bounds estimated in \cite{Ahlgren:2013wba} and the combined beam dump experiments from \cite{Williams:2011qb}. Unshaded regions represent the reach of SHIP from meson decay and bremsstrahlung searches  \cite{Gorbunov:2014wqa}. The curves indicate the values of $(g_{B-L}, m_V)$ where the lightest sterile neutrino $N_1$ can account for the whole dark matter for three values of the neutrino masses, $m_N=1,10, 100$ keV in solid, dotted and dashed lines.}
	\label{fig:SN}
\end{figure}

For $m_V \leq 1$GeV the strongest constraints come from supernova cooling \cite{Mayle:1987as,Raffelt:1987yt,Turner:1987by,Chang:2016ntp},  beam dump searches \cite{Bjorken:2009mm,Williams:2011qb,Vilain:1993kd,Bjorken:1988as,Riordan:1987aw,Bross:1989mp} and big bang nucleosynthesis (BBN)\\\cite{Williams:2011qb,Ahlgren:2013wba,Fradette:2014sza,Berger:2016vxi,Huang:2017egl}. Recent updates on these bounds are compiled in Fig.~\ref{fig:SN}. The lower mass region  labeled  BBN is excluded by   the effect of $\Delta N_{\rm eff}$ on the expansion, while the higher mass BBN region is excluded because the injection of electromagnetic energy from the  $V$ decay to charged particles during nucleosynthesis distorts the abundance of light elements. These BBN constraints have been evaluated in detail in ref. \cite{Fradette:2014sza,Berger:2016vxi}. In the relevant region of parameter space, they are seen to depend  on the lifetime of the decaying particle  and its abundance per baryon prior to decay. 
The corresponding region in Fig.~\ref{fig:SN} is a sketch of the excluded region in the latter analysis, which is approximately bounded by the lines corresponding to the lifetimes  $\tau_V \in [0.1-100]$s, the threshold for $V$ hadronic decays, $m_V \geq 2 m_\pi$, and the line corresponding to the fraction of the $V$ decaying to charged particles per baryon, $n_{V}^{em}/n_b = 1$.   
 
\begin{figure}
	\centering
		\includegraphics[width=4.cm,keepaspectratio]{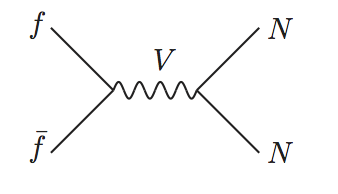} \hspace{1cm} \vspace{1cm}\includegraphics[width=2.7cm,keepaspectratio]{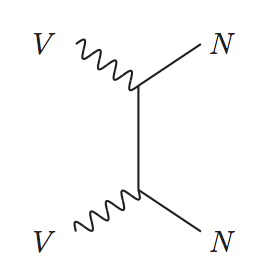}\hspace{1cm}\includegraphics[width=4.cm,keepaspectratio]{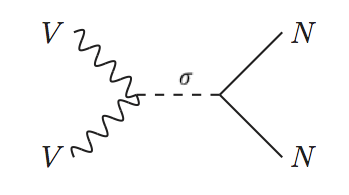}
	\caption{Production of sterile neutrinos  via the new gauge interaction.}
	\label{fig:Vnn}
\end{figure}

At least two of the neutrinos will be involved in the BO mechanism and their masses must be in the 1-100 GeV range. We need to ensure that the new $B-L$ interactions do not bring them 
to thermal equilibrium before $t_{EW}$, which will set an upper bound on $g_{B-L}$.  It it important to know however if the $B-L$ gauge boson is in thermal equilibrium which also depends on $g_{B-L}$. For $m_V \lesssim 65 \, {\rm GeV}$, the dominant process  is the scattering $ t t \leftrightarrow V H$ with a rate that can be estimated to be
\begin{equation}
	\Gamma( t t \rightarrow V H)\simeq {\zeta(3)  g_{B-L}^2 y_t^2\over 144 \pi^3}  T \log\Big(\frac{T}{m_t(T)}\Big),
	\label{thermal}
\end{equation}
where $y_t$ is the Yukawa of the top quark and $m_t(T)$ is its thermal mass. This rate is larger than the Hubble expansion somewhere above the EW transition provided $g_{B-L} \gtrsim 10^{-6}$. 
For larger masses of the gauge boson ($m_V \gtrsim {\rm 65 GeV}$) the process $ V \leftrightarrow f\bar{f}$ kinematically opens up at high temperature and one has to consider the decay and inverse decay with a rate
\begin{eqnarray}
\Gamma(V \leftrightarrow f \overline{f}) = { g_{B-L}^2 N_C Q_f^2 m_V\over 12 \pi} \left(1+ {2 m_f^2\over m_V^2}\right) \left(1-{4 m_f^2\over m_V^2} \right)^{1/2},\nonumber\\
\end{eqnarray}
where $N_C=3(1)$ and $Q_f=1/3 (-1)$ for quarks(leptons). The sum is  over all the standard model fermions whose thermal mass is such that $m_V(T) \geq 2 m_f(T)$. 
The lower limit on $g_{B-L}$ for the thermalization of the $B-L$ boson is shown as a dotted red line in Fig.\ref{fig:bound}. Provided the $B-L$ boson is in thermal equilibrium we have to consider its interactions with the sterile neutrinos driving leptogenesis. Assuming that the $V$ boson is lighter
than $2 m_N$, so that the decay $V\rightarrow NN$ is kinematically forbidden,  the dominant contribution \cite{Heeck:2016oda} comes from the scattering processes with the fermions and gauge bosons in Fig.\ref{fig:Vnn}.
 To ensure these processes do not equilibrate the sterile neutrinos, we should have
\begin{equation}
\langle \Gamma(V V \rightarrow NN) \rangle_{T=\langle\phi\rangle} < \,H(T=\langle\phi\rangle)
\label{eq:noth}
\end{equation}
Assuming $M_{\sigma} \sim \langle \phi \rangle \gg m_V, m_N$, 
\begin{eqnarray}
\langle \Gamma(VV \rightarrow NN) \rangle \sim \frac{g_{B-L}^4 m_N^2}{m_V^4}\frac{3 \zeta(3)}{4 \pi^3} T^3, 
\end{eqnarray}
and using eq.~(\ref{eq:mv}) we get
\begin{equation}
	g_{B-L} <  5\cdot 10^{-6}\Big(\frac{m_V}{1 {\rm GeV}}\Big)\Big(\frac{1{\rm GeV}}{m_N}\Big)^{2/3} \ .
	\label{naivebound}
\end{equation}
This upper bound is shown by the red solid line in Fig.\ref{fig:bound}. In the next section we will include these new interactions in the equations for the generation of the baryon asymmetry, and our results confirm the naive estimate in eq.~(\ref{naivebound}).

Finally we consider the production of neutrinos through the decay of $\sigma\rightarrow NN$, which is relevant for $T \geq M_{\sigma}$, since $M_{\sigma} \gg m_N$. The requirement that  this process does not thermalize the sterile neutrinos implies that the decay rate, $\Gamma_\sigma$,  is slower than the Hubble rate at $T\geq M_\sigma$:
\begin{equation}
\Gamma_D(M_\sigma)=\frac{h_N^2 M_{\sigma}}{16\pi}\lesssim H(M_\sigma).
\label{eq:gsigma}
\end{equation}
For $m_N = 1-100$ GeV, using eq.~(\ref{eq:mv}), we get
\begin{equation}
M_\sigma \sim \langle \phi\rangle \geq 2 \times 10^5-5 \times 10^6 {\rm GeV}.
\label{eq:faars}
\end{equation}

On the other hand, for $m_V \geq 2 m_N$ the dominant production goes via the decay of the gauge boson into two sterile neutrinos $V\rightarrow N N$, which, if kinematically allowed, scales with $g_{B-L}^2$. In this case the decay rate is 
\begin{equation}
\Gamma(V\rightarrow NN) = {g_{B-L}^2 m_V\over 24 \pi}  \left(1- {4 m_N^2\over m_V^2}\right)^{3/2}. 
\label{eq:gvnn}
\end{equation}
Requiring that it is smaller than $H(T_{EW})$ implies for $m_N \ll m_V$ 
\begin{equation}
g_{B-L} \lesssim 10^{-7}\Big(\frac{100 {\rm GeV}}{m_V}\Big)^{1\over 2}.
\label{eq:gbound}
\end{equation}

\begin{figure}
	\centering
	\includegraphics[width=8cm,keepaspectratio]{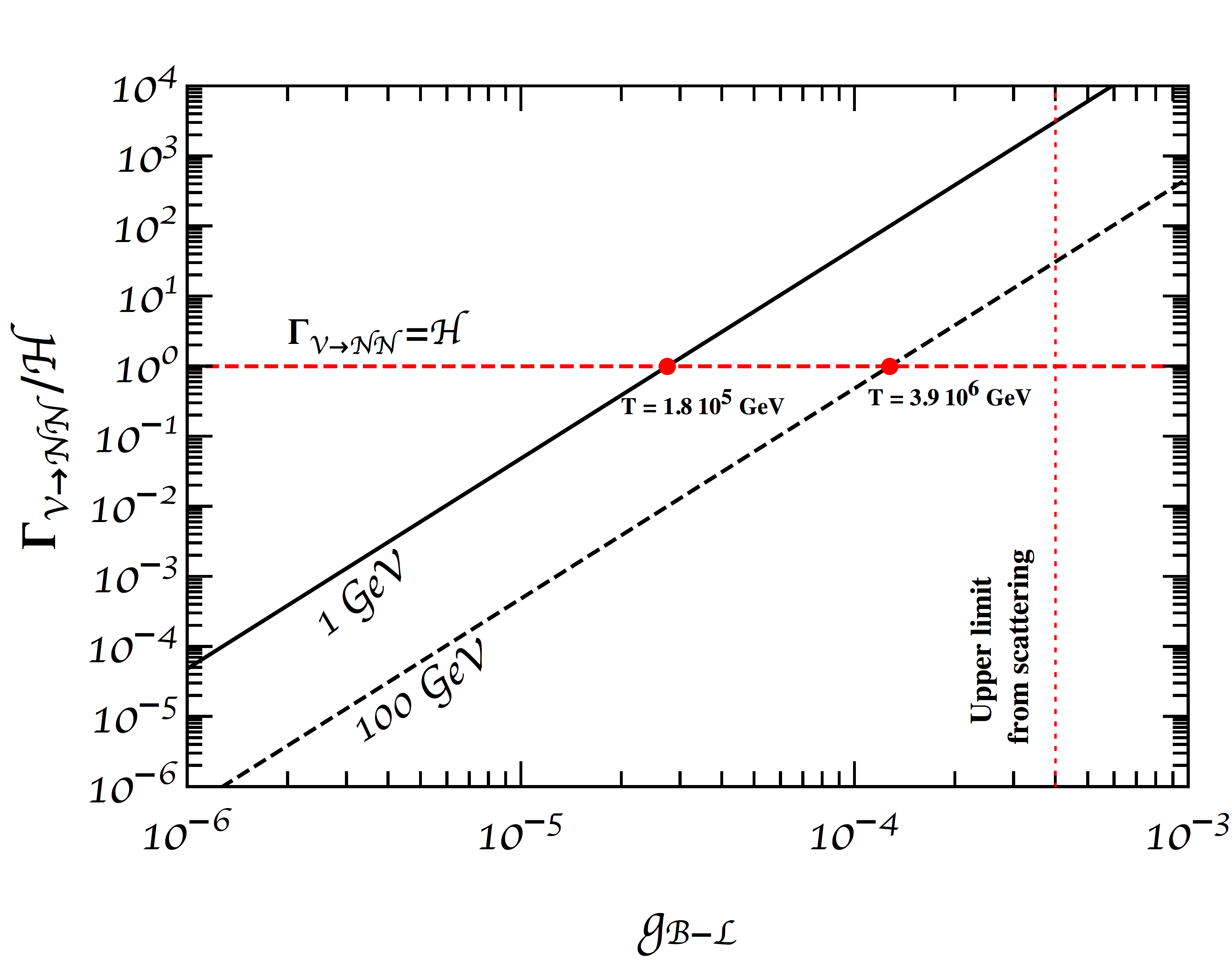}
	\caption{$\Gamma(T)/H(T)$ including thermal effects for $m_V$ and two values of the heavy neutrino masses $m_N=1, 100$GeV for $T = 2 \times$ the threshold temperature.}
	\label{fig:govh}
\end{figure}

One may worry if thermal mass corrections can allow the decay $V\rightarrow NN$ at large temperatures even if $m_V \leq 2 m_N$. At high enough temperatures both sterile neutrinos and the gauge boson acquire thermal corrections to the masses of the form
\begin{equation}
m(T)\sim  g_{B-L} T.
\end{equation}
The thermal mass of the gauge boson is  larger than that of the sterile neutrino, because all fermions  charged under $B-L$ will contribute to the former and only the gauge boson loop contributes to the later\cite{Weldon:1982bn}: \begin{equation}
m^T_V = m_V + \sqrt{{4\over 3}} g_{B-L} T, \;\;\; m^T_N = m_N + {1\over \sqrt{8}} g_{B-L} T.
\end{equation} 
We substitute the temperature dependent  mass in eq.~(\ref{eq:gvnn})  and we show  in Fig.~\ref{fig:govh} the ratio 
$\Gamma(V\rightarrow N N)/H$ close to the minimum threshold temperature (where $m_V^T \geq 2 m_N^T$),
for $m_N=1$ and 100 GeV as a function of $g_{B-L}$. The upper limit for $g_{B-L}$ are less stringent than 
 those derived  from $2\to 2$ processes in eq.~(\ref{naivebound}).We now evaluate in detail the effect on BO induced by the new scatterings of Fig.~\ref{fig:Vnn}.  Leptogenesis in the presence of a new $B-L$ gauge interaction has been recently studied in \cite{Heeck:2016oda}, although not in the context of BO, which as far as we know has not been considered before. 

\subsection{Leptogenesis}\label{lepto}

The sterile neutrinos relevant for leptogenesis are the heavier ones, $m_{N_{2,3}}$, with masses in the $1$-$100$~GeV, and we focus on the scenario where $m_V\leq 2 m_{N_{2,3}}$.   
We now explain how to include the terms involving the $B-L$ gauge interactions in the quantum kinetic equations for BO leptogenesis as derived in \cite{Hernandez:2016kel}. Following Raffelt-Sigl approach \cite{Sigl:1992fn}, we consider a density matrix, $\rho(k)$, describing the expectation value of number densities of $N$ and
a density matrix describing the corresponding anti-particles, $\bar{\rho}(k)$\footnote{We can neglect Majorana masses in the range of masses considered and therefore particles and antiparticles correspond to the two helicity states. }. These equations are complemented with three equations involving the slow varying chemical potentials, $\mu_{B/3-L_\alpha}$. The modification of the kinetic equations  induced by the $B-L$ interactions is the addition of new collision terms in the equations for $\rho$ and $\bar{\rho}$, 
that have the same flavour structure as the neutral current contribution considered in \cite{Sigl:1992fn}. As explained above the most relevant contributions come from the scattering
processes: ${\bar f} f \leftrightarrow N {\bar N}$ and $V V\leftrightarrow N {\bar N}$. The latter is enhanced at high temperatures, but of course will only be relevant if the $V's$ are in thermal equilibrium, which we assume in the following. 

The additional collision terms, from the first process, in the equation for the evolution  $\rho(k)$ can be writen in the form:
\begin{align}
({\dot{\rho}}(k))^{ff\rightarrow NN}_{B-L} ={1\over 2}  \int \prod_i {d^3 p_i\over (2\pi)^3 2 E_i} (2\pi)^4 \delta(p_1+p_2-p_3-k)  \nonumber\\
\sum_{spins} |{\mathcal M}(\bar{f}(p_1) f(p_2) \rightarrow \overline
{N}(p_3)N(k))|^2 
\nonumber\\
f^{eq}_1 f^{eq}_2 \Big[2 -\left\{r, \bar{r}\right\} - f^{eq}_4  \left\{r, 1-\bar{r}\right\} - f^{eq}_3 \left\{\bar{r}, 1-r\right\} \Big] ,
\label{eq:rhokdot}
\end{align}
where $f^{eq}_i\equiv f_F(p_i)$ is the Fermi-Dirac equilibrium distribution function of the particle with momentum $p_i$ with $p_4 \equiv k$; $\{,\}$ is the anticommutator, 
and  the normalized matrices are:
\begin{eqnarray}
r(k) \equiv {\rho(k)\over f_F(k)},\;\;\; \bar{r}(p_3) \equiv {\bar{\rho}(p_3)\over f_F(p_3)}.
\end{eqnarray}
The additional collision terms for $\bar{\rho}$ have the same form with the substitution $k\leftrightarrow p_3$ . 

For the second process we have similarly:
\begin{align}
({\dot{\rho}}(k))^{VV\rightarrow NN}_{B-L} ={1\over 2}  \int \prod_i {d^3 p_i\over (2\pi)^3 2 E_i} (2\pi)^4 \delta(p_1+p_2-p_3-k)  \nonumber\\
\sum_{spins} |{\mathcal M}(V(p_1) V(p_2) \rightarrow \overline
{N}(p_3)N(k))|^2 
\nonumber\\
f^{eq}_1 f^{eq}_2 \Big[2 -\left\{r, \bar{r}\right\} - f^{eq}_4  \left\{r, 1-\bar{r}\right\} - f^{eq}_3 \left\{\bar{r}, 1-r\right\} \Big] ,
\label{eq:rhokdotZ}
\end{align}
where $f^{eq}_i$ is the  equilibrium distribution function of the particle with momentum $p_i$, ie.  Bose-Einstein, $f^{eq}_i\equiv f_B(p_i)$, for $i=1,2$ and
Fermi-Dirac, $f^{eq}_i\equiv f_F(p_i)$, for $i=3,4$.  


As usual we are interested in the evolution in an expanding universe, where the density matrices depend on momentum, $y\equiv p/T$ and the scale factor or inverse temperature 
$x\propto T^{-1}$.  We consider the averaged momentum approximation, which assumes that all the momentum 
dependence factorizes in the Fermi-Dirac distribution and the density $r$ is just a function of the scale factor, ie.   $\rho(x,y) = f_F(y) r(x)$. In this approximation we can do the integration over momentum and the $B-L$  terms in the equation for $r$ and $\bar{r}$ become:
\begin{align}
\left( x H {d r \over d x}\right)_{B-L} = \Big( x H {d \bar{r} \over d x}\Big)_{B-L}&=& { \langle \gamma^{(0)}_V \rangle \over 2} \Big(2 -\left\{r, \bar{r}\right\}\Big)\nonumber\\
- \langle \gamma^{(1)}_V \rangle \Big(r+\bar{r} - \{r,\bar{r}\} \Big),\nonumber\\
\label{eq:rrb}
\end{align}
where $H$ is the Hubble expansion parameter.

The averaged rates including the two processes in eqs.~(\ref{eq:rhokdot}) and (\ref{eq:rhokdotZ}) (assuming in the latter the $Z's$ are in equilibrium) are computed in the appendix with the result: 
\begin{align}
\langle \gamma^{(0)}_V\rangle &=& \left(3.2(3) \times 10^{-3}+ 2.95 \times 10^{-4} {m_N^2 T^2\over m_V^4 }\right)  g_{B-L}^4 T  ,\nonumber\\
\langle \gamma^{(1)}_V\rangle &=& \left(3.4(1) \times 10^{-4} +3.55 \times 10^{-5}  {m_N^2 T^2\over m_V^4 }\right) g_{B-L}^4 T ,
\label{eq:gamv}
\end{align}
where the two terms inside the brackets correspond respectively to the $f\bar{f}$ and $VV$ channels, and are valid for $T< T_{\rm max} \equiv M_\sigma \sim \langle \phi \rangle = { m_V\over 2 \sqrt{2} g_{B-L}}$. 
Note the different temperature dependence of the two contributions. The growth of the $VV\leftrightarrow NN$ at high temperatures originates in the contribution of the longitudinal polarization of the V bosons when the temperature is below the scalar mass, $M_\sigma$.  For higher temperature, $T\geq T_{\rm max}$, 
 the contribution of the physical scalar $\sigma$ has to be included, leading to a rate $\propto T$.
The new interactions do not modify the chemical potential dependent terms, nor the evolution equation for $\mu_{B/3-L_\alpha}$. The equations are therefore those in \cite{Hernandez:2016kel} with the additional $B-L$ terms in eq.~(\ref{eq:rrb}). 

\begin{figure}
	\centering
	\includegraphics[width=7cm]{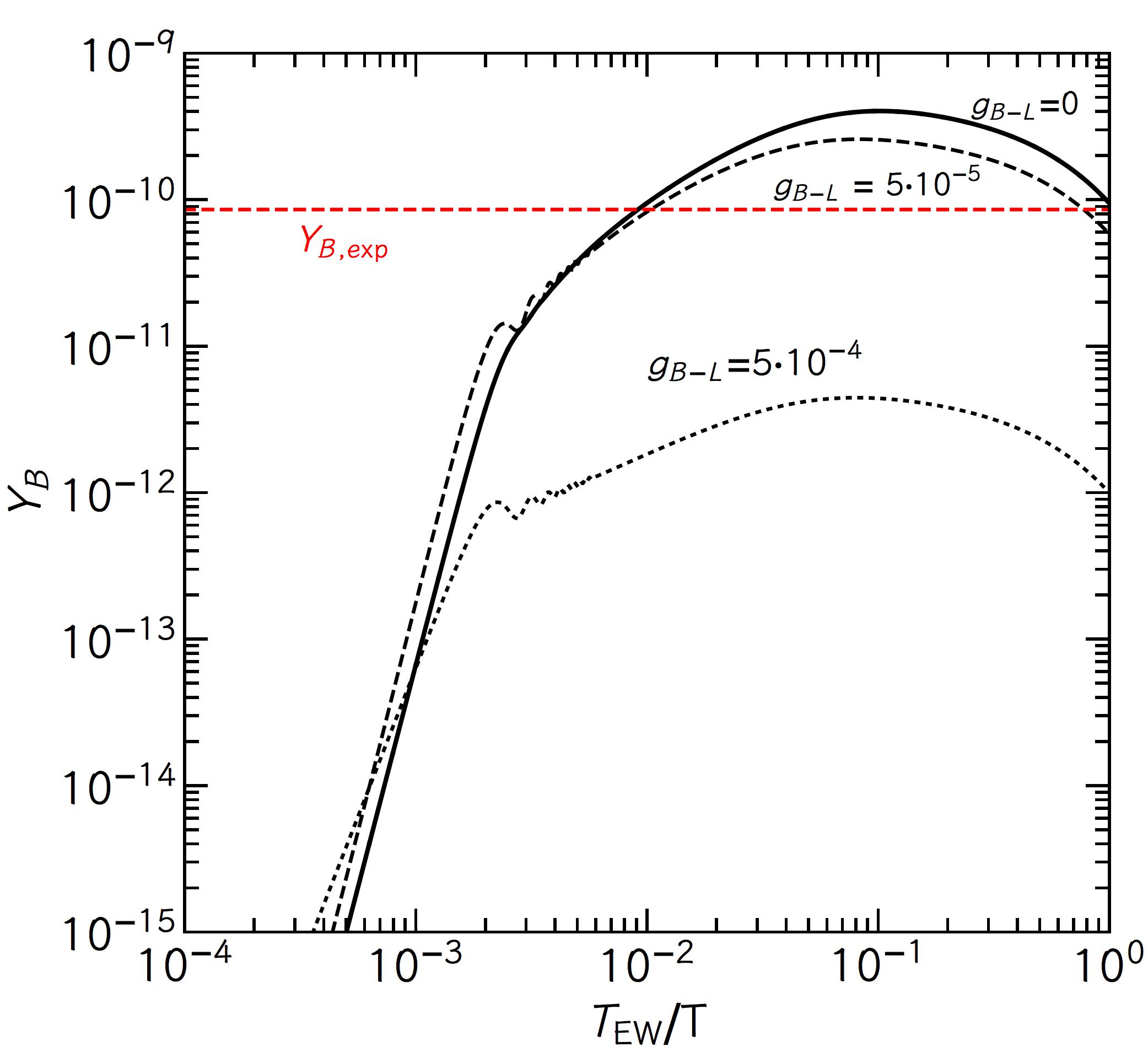}
	\caption{$Y_B$ as a function of $T_{\rm EW}/T$ for the  $Y$ and $m_N$ parameters corresponding to the test point in \cite{Hernandez:2016kel} and 
		$g_{B-L} =0 $ (solid),  $g_{B-L} = 5 \times 10^{-5}$ (dashed), $g_{B-L} = 5\times 10^{-4}$ (dotted). The horizontal line is the observed value. }
	\label{fig:YB}
\end{figure}

To illustrate the effect of the $B-L$ gauge interaction, we have considered the test point of ref.~\cite{Hernandez:2016kel} with masses for the heavy steriles $m_{N_{2,3}} \sim 0.8$ GeV. Within the parameter space of successful leptogenesis, this point was  chosen because it leads to charmed meson decays to heavy sterile neutrinos that could be observable in SHIP, and furthermore this measurement, in combination with input from  neutrinoless double beta decay and CP violation in neutrino oscillations, could provide a quantitative prediction of the baryon asymmetry. Adding the $B-L$ terms to the equations for $r$ and $\bar{r}$ of \cite{Hernandez:2016kel}, and solving them numerically  (for details on the method  see \cite{Hernandez:2016kel}) we obtain the curves in Fig.~\ref{fig:YB}.  The rates depend on $m_V$ so we choose $m_V = 1$GeV.
The evolution of the baryon asymmetry as a function of $T_{\rm EW}/T$ is shown by the solid line of Fig.~\ref{fig:YB} in the absence of $B-L$ interactions or for a sufficiently small value of $g_{B-L}$. The suppression of the asymmetry is visible for larger values of $g_{B-L}\gtrsim {\rm few} \times 10^{-5}$ as shown by the dashed and dashed-dotted lines. The naive expectations  in Fig.~{\ref{fig:bound}} is therefore confirmed and we do not expect a significant modification of the baryon asymmetry of the minimal model, as long as $g_{B-L}$ satisfies the bound in eq.~(\ref{naivebound}).

\subsection{Dark Matter}

Now we want to discuss possible dark matter candidates in the $B-L$ scenario without spoiling the BO mechanism, which as we have seen imposes a stringent upper bound on the gauge 
coupling, $g_{B-L}$. We will be interested in the region where the $V$ boson can decay to the lightest neutrino, ie $m_V \geq 2 m_{N_1}$.  The small value needed for $g_{B-L}$ suggests to consider the possibility of  a freeze-in scenario \cite{McDonald:2001vt,Hall:2009bx}, where the gauge boson does not reach thermalization, and neither does the lightest sterile neutrino, $N_1$. The status of dark matter in a higher mass range through freeze-out has been recently updated in \cite{Escudero:2018fwn}.

As it is well known, $N_1$ in the keV mass range is sufficiently long lived to provide a viable warm DM candidate \cite{Dodelson:1993je,Shi:1998km}. The $B-L$ model is as we will see a simple extension of the 
$\nu$MSM \cite{Asaka:2005pn}, which avoids the need of huge lepton asymmetries to  evade X-ray bounds. In our scenario the keV state is produced from the decay $V\rightarrow N_1 N_1$, while the lifetime of $N_1$, relevant in X-ray bounds, is controlled also by mixing, which can be sufficiently small, in a technically natural way, as in the $\nu$MSM scenario (provided the lightest neutrino mass is small enough). A similar scenario for DM has been studied in \cite{Kaneta:2016vkq}. We now quantify the parameter space for successful DM and leptogenesis in this scenario.

We assume that the abundance of $V$ and $N_1$ is zero at a temperature below the EW phase transition where all the remaining particles in the model are in thermal
equilibrium. All fermions in the model couple to the $V$ and therefore its production is dominated by the inverse decay process: $f \overline{f} \rightarrow V$. The kinetic equation describing the production of $V$ is the following:
\begin{equation}\label{BoltzmannV}
\begin{split}
\dot{n}_V+3Hn_V=\sum_f \int\frac{d^3p_f}{(2\pi)^3 2E_f}\frac{d^3p_{\bar{f}}}{(2\pi)^3 2E_{\bar{f}}}\frac{d^3p_V}{(2\pi)^3 2E_V}\\(2\pi)^4\delta^{4}(p_V-p_f-p_{\bar{f}})[|M|^2_{f\bar{f}\rightarrow V}f_f f_{\bar{f}}(1+f_V)+\\-|M|^2_{V\rightarrow f\bar{f}}f_V(1- f_f)(1- f_{\bar{f}})],
\end{split}
\end{equation}
where $f_i(p)$ are the distribution function of the particle, $i$, with momentum $p$, and 
\begin{equation}
n_i = g_i \int {d^3 p\over (2 \pi)^3} f_i(p),
\end{equation}
is the number density, with $g_i$ the number of spin degrees of freedom. $g_f=4$ for a Dirac fermion, $g_N=2$ for a Majorana fermion and $g_V=3$ for a massive gauge boson.   $M$ is the amplitude for the decay $V\rightarrow f \overline{f}$ at tree level. 

The sum over $f$ is over all fermions, but we can safely neglect the contribution of the $N_1$ and also those that are non-relativistic.  We can also neglect the Pauli-blocking and stimulated emission effects ($f_i\pm1 \sim \pm 1$) and approximate the distribution function in equilibrium for fermions and bosons by the Maxwell-Boltzmann, $f_i(p_i) = f^{eq}(p_i)= e^{-E_i/T}$. Taking into account the relation
\begin{equation}
|M|^2_{V\rightarrow f\bar{f}}=|M|^2_{f\bar{f}\rightarrow V},
\end{equation}%
and the principle of detailed balance 
\begin{equation}
f_f^{eq} f_{\bar{f}}^{eq}=f_V^{eq},
\end{equation}
the equation can be simplified to
\begin{equation}
\begin{split}
\dot{n}_V+3Hn_V=-\sum_f \int\frac{d^3p_f}{(2\pi)^3 2E_f}\frac{d^3p_{\bar{f}}}{(2\pi)^3 2E_{\bar{f}}}\frac{d^3p_V}{(2\pi)^3 2E_V}\\(2\pi)^4\delta^{4}(p_V-p_f-p_{\bar{f}})|M|^2_{V\rightarrow f\bar{f}}\left( f_V(p_V)-f^{eq}(p_V)\right).
\end{split}
\end{equation}
As long as $f_V \ll f^{eq}$, the first term on the right-hand side can be neglected and the equation simplifies further to:
\begin{equation}
\dot{n}_V+3Hn_V\simeq 3 \sum_f\frac{m_V^2\Gamma_{V\rightarrow f \bar{f}}}{2\pi^2}TK_1\left(\frac{m_V}{T}\right),
\end{equation}
where $K_1$ is the first modified Bessel Function of the 2nd kind. The decay width in the $V$ rest frame is given by
\begin{eqnarray}
\Gamma(V\rightarrow f \overline{f}) = { g_{B-L}^2 N_C Q_f^2 m_V\over 12 \pi} \left(1+ {2 m_f^2\over m_V^2}\right) \left(1-{4 m_f^2\over m_V^2} \right)^{1/2},\nonumber\\
\end{eqnarray}
where $N_C=3(1)$ and $Q_f=1/3 (-1)$ for quarks(leptons). 

As usual we  define the yield of particle $i$ as
\begin{eqnarray}
Y_i = {n_i \over s},
\end{eqnarray}
where $s$ is the entropy density
\begin{eqnarray}
s= {2 \pi^2 \over 45} g^*_s T^3,
\end{eqnarray}
and we can assume $g^*_s \simeq g^*$. We also consider the averaged momentum approximation which amounts to assuming that $f_V$ has the same momentum dependence as $f^{eq}$. Changing variable from time to temperature, the final evolution equation for $Y_V$ reads:
\begin{eqnarray}
{d Y_V\over d T} = -3 \sum_f\frac{m_V^2\Gamma_{V\rightarrow f \bar{f}}}{2\pi^2 H s \left[1+ {1\over 3} {T d {g}_*\over g_* d T }\right] }K_1\left(\frac{m_V}{T}\right) \left( 1- {s Y_V   \over n_V^{eq}}\right),\nonumber\\
\end{eqnarray}
where $n_V^{eq} = {3\over 2\pi^2} m_V^2 T K_2(m_V/T)$. 

\begin{figure}
	\centering
	\includegraphics[height=7cm]{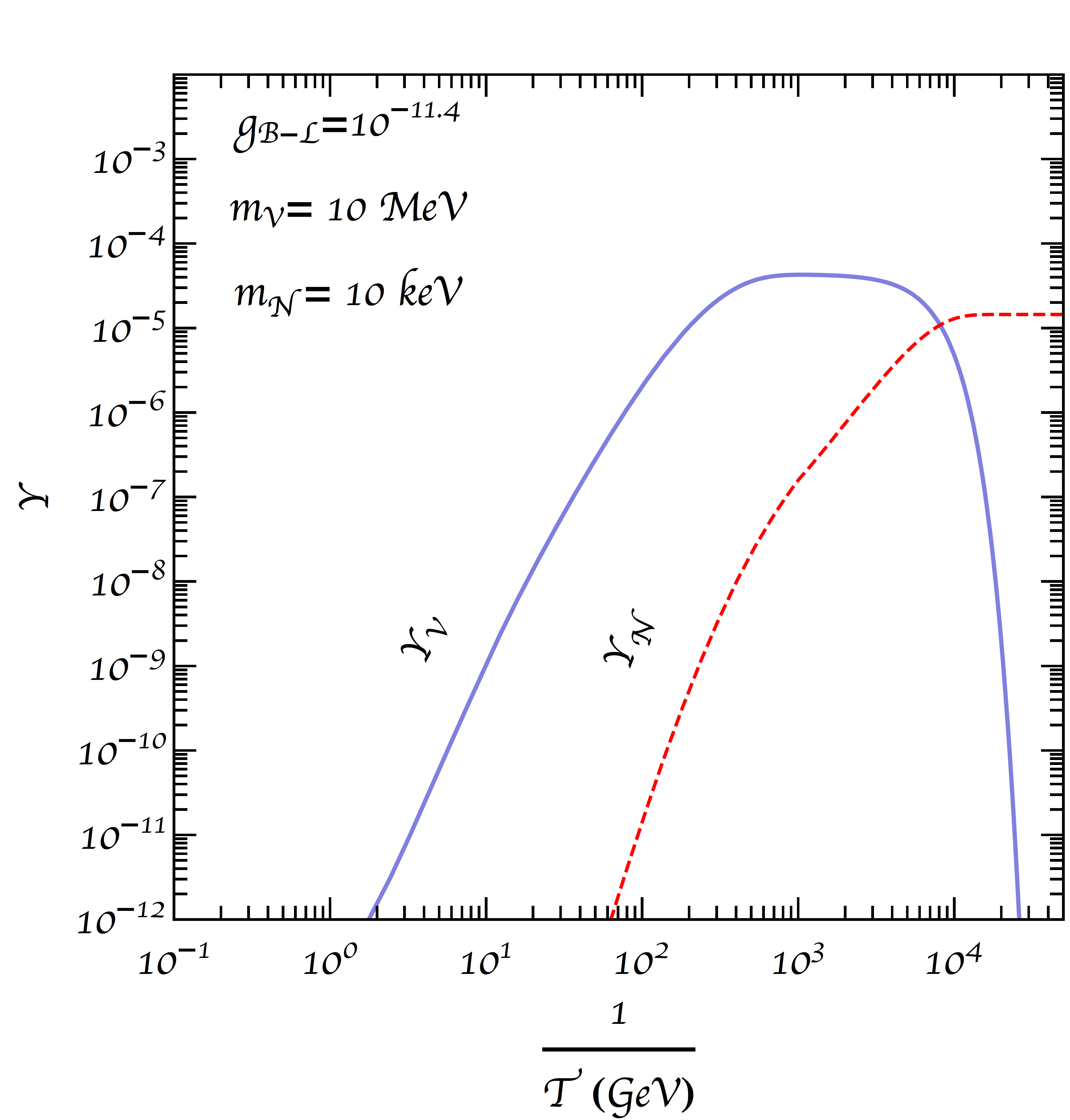}
	\caption{$V$ and $N_1$ yields as a function of the inverse temperature for $m_V= 10$ MeV and $g_{B-L} = 10^{-11.4}$.  }
	\label{fig:Ys}
\end{figure}

\begin{figure}
	\centering
	\includegraphics[width=7.cm,keepaspectratio]{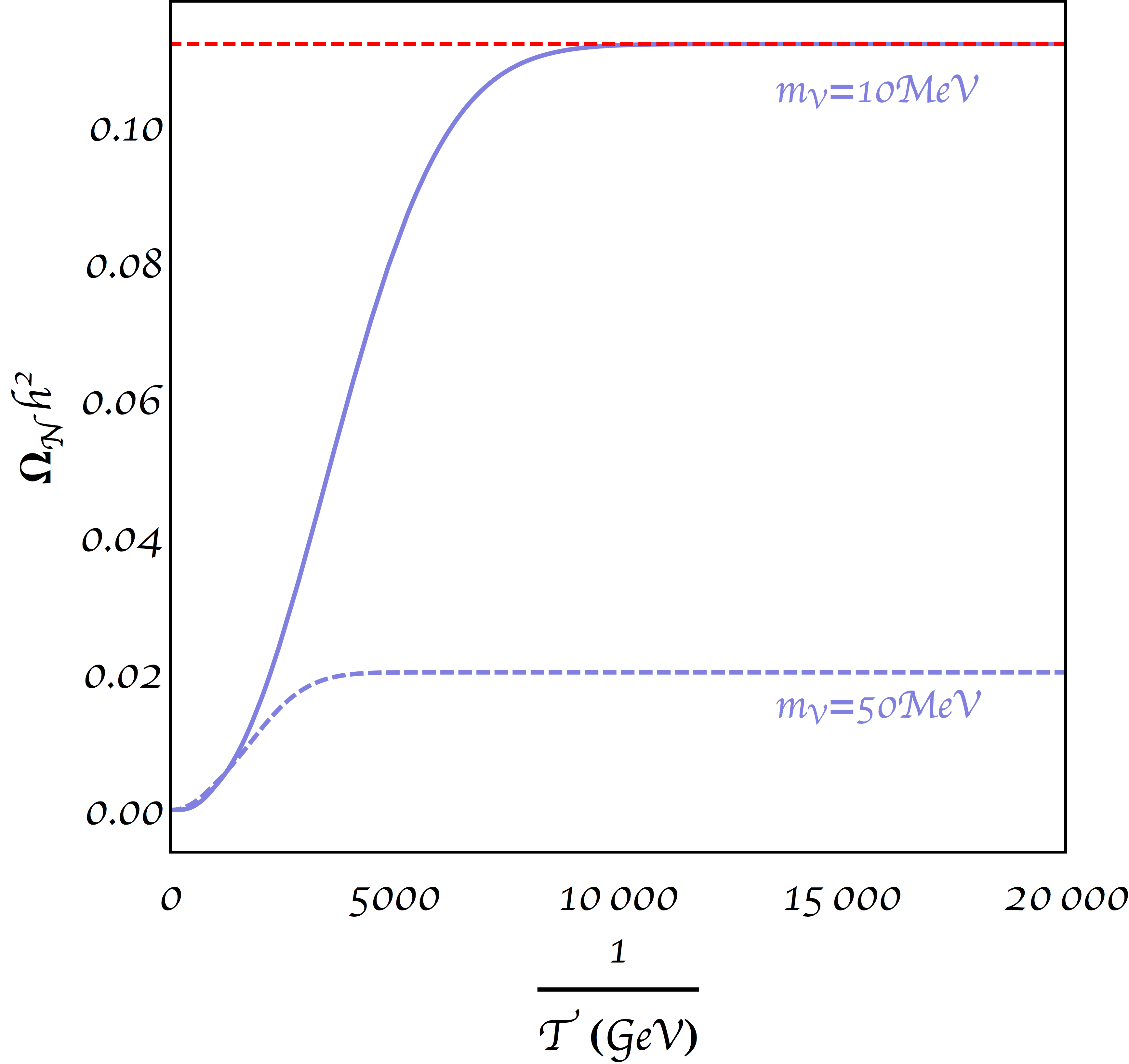}
	\caption{The evolution of the density $\Omega_Nh^2$ of the sterile neutrino $N_1$ in the $B-L$ model as a function of $1/T(GeV)$. Here we have fixed $m_{N_1}=10$ keV, $g_{B-L}=10^{-11.4}$, and used two different value of the boson mass: $m_V=10, 50$ MeV. The gray line indicates the experimental value for dark matter abundance today.}
	\label{fig:MvChangeMk8}
\end{figure}

The production of $N_1$ is dominated by the decay $V\rightarrow N_1 N_1$. There is also the contribution via mixing with the active neutrinos but this is negligible for mixings that evade present X ray bounds. Neglecting the inverse processes, the evolution equation for $n_1$ is
\begin{equation}
\dot{n}_{1}+3Hn_{1}=2\frac{K_1(x)}{K_2(x)}\Gamma(V\rightarrow N_1 N_1) n_V,
\end{equation}
and in terms of the yield
\begin{eqnarray}
{d Y_{N_1}\over d T}  = - {2 \over H T  \left[1+ {1\over 3} {T d {g}_*\over g_* d T }\right] }  \frac{K_1(x)}{K_2(x)} \Gamma(V\rightarrow N_1 N_1) Y_V,\nonumber\\
\end{eqnarray}
where
\begin{eqnarray}
\Gamma(V\rightarrow N_1 N_1) = { g_{B-L}^2  m_V\over 24 \pi} \left(1-{4 m_{N_1}^2\over m_V^2} \right)^{3/2}.
\end{eqnarray}
It is straightforward to solve these equations. In Fig.~\ref{fig:Ys} we show the yields of $V$ and $N$ as function of the inverse temperature for $m_V =10$ MeV, $m_{N_1}= 10$ keV and $g_{B-L} =10^{-11.4} $. 
The resulting  abundance of $N_1$ is
\begin{eqnarray}
\Omega_{N_1} h^2 \equiv {s_0 m_{N_1} \over \rho_c h^{-2}} Y_{N_1} \simeq 2.7 \times 10^2 Y_{N_1} {m_{N_1}\over {\rm keV} }, 
\end{eqnarray}
where $s_0$ = 2889.2 cm$^{-3}$ is the entropy today and $\rho_c= 1.05 10^{-5} h^2$ GeV cm$^{-3}$ is the critical density. The evolution of $\Omega_{N_1} h^2$  is shown in 
Fig.\ref{fig:MvChangeMk8} for two values of $m_V$ and a fixed value of $g_{B-L}$. 
Requiring that $\Omega_{N_1} h^2$ equals the full DM contribution of 
$\Omega_{DM} h^2 \simeq 0.12$ implies a relation between $m_V$ and $g_{B-L}$ as shown in the curves of Fig.~\ref{fig:SN}.The values of $g_{B-L}$ corresponding to the right dark matter relic abundance do not affect leptogenesis and lie far below the actual collider limits. Nevertheless some regions of the parameter space are interestingly excluded from supernova and BBN observations. 

 A final comment concerns the comparison of our calculation of the DM abundance and that in ref.~\cite{Kaneta:2016vkq}. In this reference only the evolution of the $N_1$ is considered, and the collision term corresponds to the scattering process $f \bar{f} \rightarrow N_1 N_1$, where the narrow width approximation is assumed. We believe this method is only equivalent to ours when all $f, \bar{f}$ and $V$ distributions are the equilibrium ones, but this is not the case here.   In the region they can be compared our results are roughly a factor three smaller than those in \cite{Kaneta:2016vkq}. 

\subsection{Couplings}\label{masses}
According to the previous calculation, the relic DM abundance requires a very small $B-L$ coupling. 
In order to obtain, for example, a mass $m_V\sim 1$ MeV, the gauge coupling needed to generate DM  is
\begin{equation}
g_{B-L}\sim 10^{-11.8} ,
\end{equation}
and therefore 
\begin{equation}
\langle \phi\rangle \sim 2 \cdot10^8 \, {\rm GeV}.
\label{eq:avephi}
\end{equation}

In order to get the $N_1$ and $N_{2,3}$ in the target range of keV and 1-100 GeV respectively, small and hierarchical $h_{N_i}$ couplings are needed:
\begin{equation} 
h_{N_{2}}\simeq h_{N_3} \sim10^{-6}-10^{-8},
\end{equation}
for the heavy sterile neutrinos involved in BO leptogenesis and 
\begin{equation}
h_{N_1}\sim10^{-12},
\end{equation} 
for the dark matter candidate. Note that the required $h_N$'s couplings are in the same ballpark as the yukawa couplings. 
The gauged $B-L$ model works nicely to explain neutrino masses, the baryon asymmetry and dark matter. Unfortunately it also requires a very small $g_{B-L}$ which will be
very hard to test experimentally. An alternative might be to consider a flavoured $U(1)$, for example $L_\mu -L_\tau$, that might be compatible with a larger $g_{B-L}$, provided the assignment of charges to the singlet states
ensures that not all of them reach thermalization via the flavoured gauge interaction before $t_{EW}$.
\section{Axion and Neutrinos}\label{axion}

As a second example we consider an extension of eq.~(\ref{eq:seesaw}) with a scalar doublet and a scalar singlet. This model is also an extension of the invisible axion model \cite{Dine:1982ah} with  sterile neutrinos, that was first considered in  \cite{Langacker:1986rj}, providing a connection between the Peccei-Quinn (PQ) symmetry breaking scale and the seesaw scale
of the neutrino masses. 
The model contains two scalar doublets, $\Phi_i$, and one singlet, $\phi$. A $U(1)_{PQ}$ global symmetry exists if the  two Higgs doublets couple separately to the up and down quarks
and leptons so that the Yukawa Lagrangian takes the form:
\begin{align}\label{Eq.3}
\Lagr \supset -Y_{u}\overline{Q}_L \Phi_1 u_R-Y_{d}\overline{Q}_L \Phi_2 d_R -Y \overline{L}_L \Phi_1 N \nonumber\\-Y_l\overline{L}_L \Phi_2 l_R- {h_N\over \sqrt{2}}\overline{N^c} N \phi+ h.c. \end{align}
leading naturally to type II two-Higgs-doublet models without FCNC \cite{Branco:2011iw,Espriu:2015mfa}.

The most general scalar potential of the model compatible with a global $U(1)_{PQ}$ is the following

\begin{eqnarray}
V&=&m_1^2 |\Phi_1|^2+m_2^2 |\Phi_2|^2+m^2|\phi|^2\nonumber\\
&+&\frac{\lambda_1}{2}|\Phi_1|^4+\frac{\lambda_2}{2}|\Phi_2|^4+ \lambda_3|\Phi_1|^2 |\Phi_2|^2+\lambda_4(\Phi_1^{\dagger}\Phi_2)(\Phi_2^{\dagger}\Phi_1) \nonumber\\
& +& \frac{\lambda_\phi}{2} |\phi|^4+ \lambda_{1\phi} |\Phi_1|^2 |\phi |^2 + \lambda_{2\phi} |\Phi_2|^2 |\phi |^2 + k ( \Phi_1^\dagger \Phi_2) \phi^2.
\label{vpq}
\end{eqnarray}
The couplings in this potential can be chosen such that $\phi$ gets an expectation value,  
\begin{equation}
\left \langle \phi \right \rangle=\frac{1}{\sqrt{2} }f_{a},
\end{equation}
$U(1)_{PQ}$ is then spontaneously broken and a Nambu-Goldstone boson appears, the QCD axion. Furthermore the Majorana singlets $N$ get a mass. Expanding around the right vacuum, the field 
can be writen as
\begin{equation}
\phi=\frac{1}{\sqrt{2}}(f_a+\sigma+i a),
\end{equation}
where $\sigma$ is a massive field, while $a$ is the axion. Therefore after symmetry breaking we obtain an interaction term between sterile neutrinos and axions
\begin{equation}
\Lagr \supset -\frac{i h_N}{2} a \overline{N^c} N + h.c.
\end{equation}
The breaking scale $f_a$ must be much larger than the vacuum expectation values of the doublets,  $\gg v_{1,2}$, so that the axion can evade the stringent bounds from rare meson decays and supernova cooling, which sets a stringent lower bound $f_a \geq 4\cdot10^8$GeV \cite{Raffelt:1999tx}.

The mass of the axion is induced  by the QCD anomaly in the sub-eV range:
\begin{equation}
m_a\simeq {z^{1/2}\over 1+z} \frac{m_{\pi} f_{\pi}}{\left \langle \phi \right \rangle},
\end{equation}
where $z=m_u/m_d$. For $f_a \geq 4 \cdot 10^8$ GeV, we have
\begin{equation}
m_a\leq  {\mathcal O}\left(10^{-2}\right) {\rm eV}.
\end{equation}

\begin{figure}
	\centering
	\includegraphics[height=6
	cm,keepaspectratio]{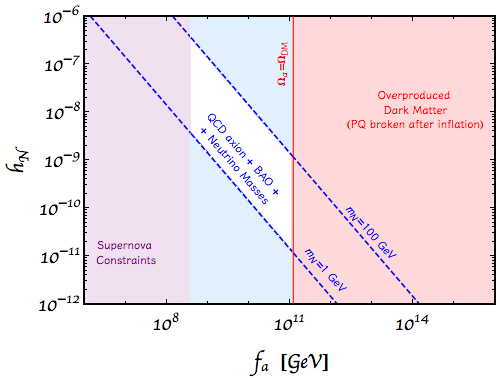}
	\caption{The light purple region is forbidden by supernova cooling constraints (left band) and the red one by axions over closing the universe (right band). Axions can explain the DM relic density in the vertical line, if inflation happens before PQ symmetry breaking. Within the remaining parameter space, in the region between the two dashed lines successful  leptogenesis  via oscillations is possible. }
	\label{fig:hmsigma}
\end{figure}

It is well known that the invisible axion is a viable cold DM candidate, through the misalignment mechanism \cite{Preskill:1982cy,Dine:1982ah,Abbott:1982af} (for recent reviews see \cite{Kawasaki:2013ae,Marsh:2015xka}).
The DM energy density is given by 
\begin{equation}
\Omega_a h^2\sim2\cdot10^{4} \left(\frac{f_a}{10^{16} {\rm GeV}}\right)^{7/6}\langle \theta_0\rangle^2,
\end{equation}
where $\theta_0$ is the misalignment angle. 
The constraints on $f_a$ depend on whether  the breaking of the PQ symmetry  happens before or after inflation; in the latter case the misalignment angle can be averaged  over many patches
\begin{equation}
\langle \theta_0\rangle^2\sim \frac{\pi^2}{3},
\end{equation}
so $\Omega_a \leq \Omega_{\rm DM}$ implies
\begin{equation}
f_a\lesssim 1.2\cdot10^{11}{\rm GeV},
\end{equation}
with the equality reproducing the observed cold dark matter energy density $\Omega_{\rm CDM}h^2\sim 0.12$. This correspond to the solid line in Fig.~\ref{fig:hmsigma}. If the PQ 
symmetry is broken before inflation, $\theta_0$ is a free parameter and the value of $f_a$ to account for DM is inversely proportional to $\theta_0^2$. 


The axion can also manifest itself as dark radiation \cite{Weinberg:2013kea}, given that it is also thermally produced \cite{Graf:2010tv}. This population of hot axions contributes to the effective number of relativistic species, but the size of this contribution is currently well within the observational bounds \cite{Salvio:2013iaa}.

In this model the VEV of the scalar singlet gives a Majorana mass to the sterile neutrinos:
\begin{equation}
m_N\approx h_N f_a.
\end{equation}
So, if we want a mass in the electroweak range, ${\mathcal O}(1-10^2)$ GeV
and $f_a\in [10^{8},10^{11}]$ GeV,    we need the  coupling  $h_N$ to be in the range:
\begin{equation}
h_N\in[10^{-11},10^{-6}]. 
\label{eq:hnrange}
\end{equation}
The hierarchy between $f_a$ and the electroweak scale requires that some couplings  in the scalar potential in  eq.~(\ref{vpq}) $(k, \lambda_{1\phi}, \lambda_{2\phi})$ are very small.  Even if not very appealing theoretically,  these small numbers are technically natural as already pointed out in \cite{Clarke:2015bea}, where the authors studied the same model with very heavy sterile neutrinos.\\
A relevant question is that of naturalness or fine-tunning of the Higgs mass in this model. In \cite{Clarke:2015bea}, this issue was studied in  the context of high-scale thermal leptogenesis, and it was concluded that stability imposes relevant constraints. In particular, a relatively small $v_2\lesssim 30$ GeV is necessary to ensure viable leptogenesis  for lower $m_N = 10^5-10^6$GeV so that yukawa's are small enough, $y \leq 10^{-4}$, and do not  induce unnaturally large corrections to the Higgs mass. 
In our case, the yukawa couplings, eq.(\ref{eq:y}), are too small to give large corrections to the Higgs mass , so no additional constraint needs to be imposed on $v_2$. As a consequence other invisible axion models, such as the KSVZ \cite{Kim:1979if,Shifman:1979if}, would also work in the context of low-scale $m_N$, but 
leads to tension with stability bounds in  the high-scale version \cite{Ballesteros:2016euj,Ballesteros:2016xej}. 

\subsection{Baryon Asymmetry}

The possibility to generate the baryon asymmetry in this model a la Fukugita-Yanagida for very heavy neutrinos $m_N \sim f_a$ was recognized in the original proposal \cite{Langacker:1986rj} and further elaborated in \cite{Clarke:2015bea}. We want to point out here that for much smaller values of $h_N$, the BO mechanism could also work successfully. 
\\As explained above the crucial point is whether the new interactions of the sterile states in this model are fast enough to equilibrate all the sterile neutrinos before EW phase transition. 
The leading order process we have to consider is the decay of the scalar into two sterile neutrinos $\sigma \rightarrow N N$, exactly as we considered in the previous section.  The limit of $M_{\sigma}\sim f_a \geq  2\times 10^5-5\times 10^6$ GeV derived in eq.~(\ref{eq:faars}) also applies here, which is safely satisfied given the supernova cooling bounds. 

\begin{figure}
	\centering
	\includegraphics[width=9cm]{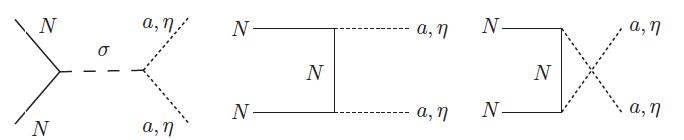}
	\caption{Annihilations of sterile neutrinos into Majorons.}
	\label{fig:NNMajoron}
\end{figure}

At second order, we must also consider the new annihilation process of sterile neutrinos to axions $N N \leftrightarrow a a$ as shown in Fig.~\ref{fig:NNMajoron}.  The rate of this process at high temperatures, $T \gg m_N$, is given by
\begin{equation}
\Gamma_{N_a}= \frac{T^3 m_N^2}{192\pi f_a^4}.
\end{equation}
The condition $\Gamma_{N_a}(T)<H(T)$ is satisfied for $T\leq f_a$ if 
\begin{equation}
f_a \geq 1.2\cdot10^{5} \left(\frac{m_N}{1 {\rm GeV}}\right)^{2/3} {\rm GeV} 
\label{eq:flimit}
\end{equation}
(for $m_N \in [1,10^2]$ GeV), safely within the targeted range. 
Fig.\ref{fig:hmsigma} shows the region on the $(f_a,h_N)$ plane  for which  successful baryogenesis  through the BO mechanism and DM can work in this model.  

Even if the necessary condition for BO leptogenesis is met for $f_a \geq 10^8$ GeV, the presence of the extra degrees of freedom, the axion, the heavy scalar and the second doublet could modify quantitatively the baryon asymmetry. For example,  the presence of two scalar doublets could modify the scattering rates of the sterile neutrinos considered in the BO scenario, where the main contributions 
\cite{Asaka:2006rw,Besak:2012qm,Ghisoiu:2014ena,Ghiglieri:2017gjz} are:
\begin{itemize}
	\item $2\leftrightarrow 2$ scatterings on top quarks via higgs exchange
	\item $2\leftrightarrow 2$ scatterings on gauge bosons 
	\item $1 \leftrightarrow 2$ decays or inverse decays including resumed soft-gauge interactions 
\end{itemize}
Sterile neutrinos are coupled to the same Higgs doublet that also couples to the top quarks; in this case nothing changes with respect to the usual calculation, in which the reactions with top quarks are mediated by $\Phi_1$. 
However, an alternative model, with a different $U(1)_{PQ}$ charge assignment, is also possible, in which the sterile neutrinos couple to $\phi_2$ and not $\phi_1$, as done in 
\cite{Clarke:2015bea}.
 In this case top quark scattering does not contribute to sterile neutrino production at tree level, but the baryon asymmetry is not expected to change significantly, since the scattering rate on gauge bosons and the 1 $\rightarrow$ 2 processes are equally important  \cite{Besak:2012qm,Ghisoiu:2014ena}. 
\\The process $\sigma \rightarrow NN$ is not foreseen to be relevant for $M_\sigma \sim f_a$, since the scalar is long decoupled when the generation of the asymmetry starts, while the new process $NN \leftrightarrow a a$ is expected to be very small according to the above estimates. It could nevertheless be interesting to look for possible corners of parameter space where the differences with respect to the minimal model is not negligible since this could provide a testing ground for the axion sector of the model.

\section{Majoron model}\label{majoron}

In between the two models described in \ref{B-L} - \ref{axion}, there is the possibility of having a global $U(1)$ spontaneously broken, which is not related to the strong CP problem and we call it lepton number. This is of course the well-known singlet majoron model \cite{Chikashige:1980ui,Schechter:1981cv,Cline:1993ht}. We assume the sterile neutrinos carry lepton number, $L_N=1$, but Majorana masses are forbidden and replaced by a yukawa interaction as in the $B-L$ model:
\begin{equation}
{\mathcal L} \supset - \left(\overline{L} Y N \Phi +\frac{1}{\sqrt{2}}h_N\overline{N^c} N \phi +h.c.\right)
\end{equation}
where $\Phi$ is the standard model Higgs doublet, while $\phi$ is a complex scalar  which carries lepton number $L_{\phi}=-2$. Then, the complex scalar acquires a VEV
\begin{equation}
\phi=\frac{f+\sigma+i\eta}{\sqrt{2}}
\end{equation}
and the $U(1)_L$ is  spontaneously broken giving rise to the right-handed Majorana mass matrix and leading to a Goldstone boson $\eta$, the majoron. Consequently the Lagrangian will induce the new scattering processes for neutrinos depicted in Fig.~\ref{fig:NNMajoron}.

As usual we have to ensure that at least one sterile neutrino does not equilibrate before $T_{EW}$ (see \cite{Gu:2009hn,Dev:2017xry} for a recent discussion in the standard high-scale  or resonant leptogenesis). As in the previous cases we have to consider the decay $\sigma\rightarrow N_{i}N_{i}$ and the annihilation into majorons, Fig.~\ref{fig:NNMajoron}. The former gives the strongest constraint, 
as in eq.~(\ref{eq:faars}):
\begin{equation}
M_\sigma \sim f \geq 2 \times 10^5-5 \times 10^6 {\rm GeV}.
\label{eq:fmaj}
\end{equation}
These lower  bounds for $m_N=1$ and 100 GeV are  shown by the horizontal lines in  Fig.~\ref{fig:majorondm}. 

\subsection{Dark Matter}

There are two candidates in this model for dark matter that we consider in turn: the Majoron and the lightest sterile neutrino, $N_1$. 

\subsubsection{Majoron }

In this model a natural candidate for dark matter is the majoron itself, but it has to acquire a mass, therefore becoming a pseudo Nambu-Goldstone boson (pNGB). One possibility is to appeal to gravitational effects \cite{Akhmedov:1992hi,Rothstein:1992rh}. However,  the contribution to the mass from gravitational instantons  is estimated to be \cite{Hebecker:2016dsw,Alonso:2017avz,Lee:1988ge}
\begin{equation}
m_\eta \sim M_{P} e^{-\frac{M_{P}}{f}},
\end{equation}
and therefore extremely tiny, unless $f$ is close to the Planck scale. 

Another alternative is to consider a flavoured $U(1)_X$ and  soft symmetry breaking terms in the form of yukawa couplings \cite{Hill:1988bu,Frigerio:2011in}. This possibility has been studied in detail in ref.~ \cite{Frigerio:2011in}. It has been shown that the majoron can be the main component of dark matter for sterile neutrino masses $m_N \geq 10^5$ GeV, while for masses in the range we are interested in ($m_N\sim 1-100$ GeV) neither thermal production via freeze-out nor via freeze-in works.  

The possibility  to produce it via vacuum misalignment, analogous to the one which produces the axion relic density has also been discussed in \cite{Frigerio:2011in}. It was  shown  to give a negligible contribution compared to the thermal one, because the majoron gets a temperature dependent mass at early times. Even if the mass of the majoron is significantly smaller 
in our situation, with lighter $m_N$,  we find the same result, ie. that only a small fraction of the DM can be produced via misalignmet.  

\begin{figure}
	\centering
	\includegraphics[height=7cm,keepaspectratio]{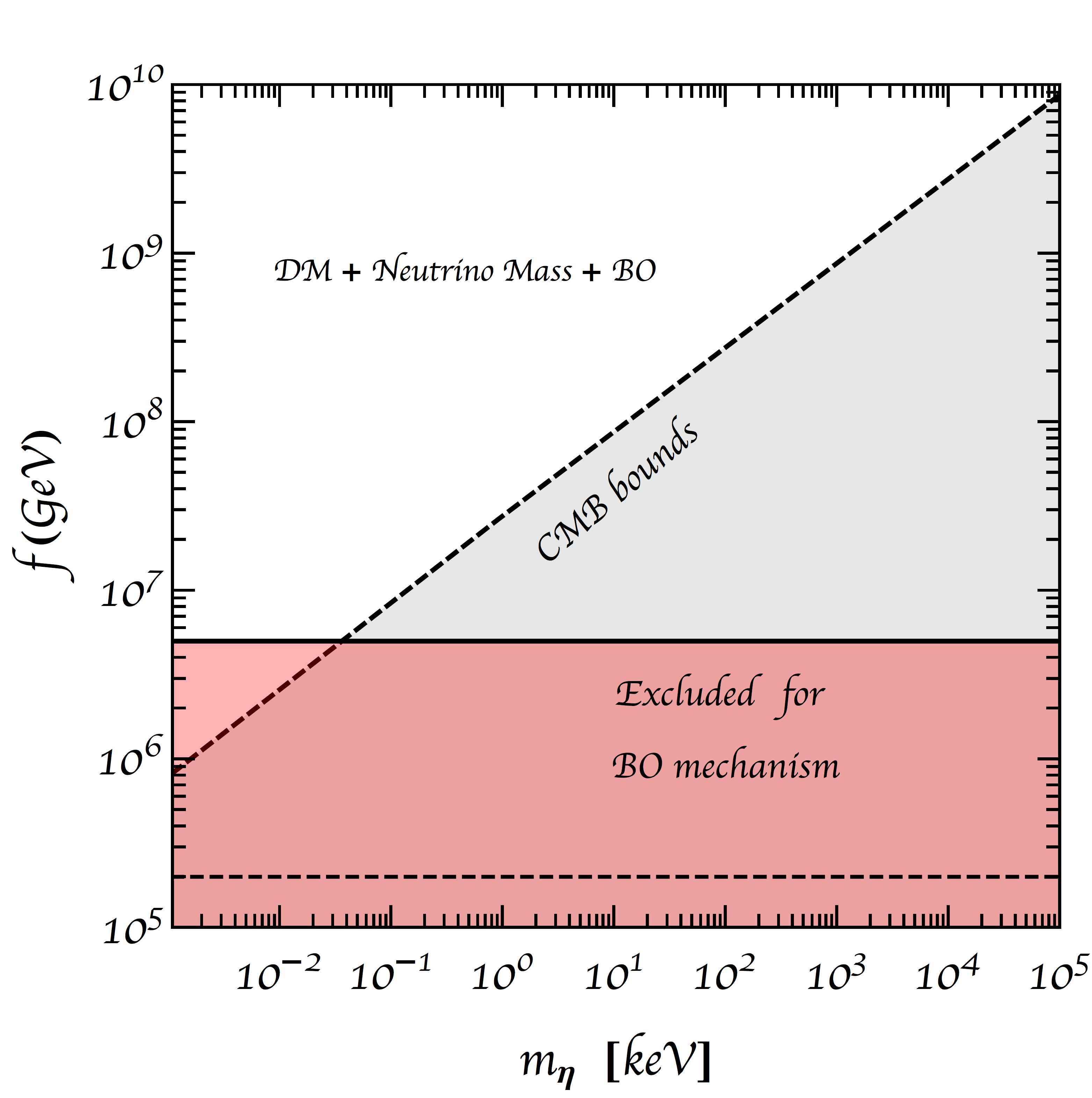}
	\caption{Constraints in the plane $f-m_{\eta}$ considering the Majoron as the only dark matter component. The region in gray below the dashed line is excluded from CMB measurements\cite{Lattanzi:2014mia}, while the light red region below the solid (dashed) line is excluded in order not to spoil ARS for $M_{N} =$ 100 (1) GeV.}
	\label{fig:majorondm}
\end{figure}

No matter what the production mechanism is  if the majoron constitutes the dark matter, there are constraints from the requirement that the majoron be stable on a cosmological timescale and its decay 
to the light neutrinos 
\begin{equation}
\Gamma(\eta \rightarrow \nu\nu)=\frac{1}{64\pi}\frac{\sum_i m^2_{\nu_i}}{f^2}m_{\eta}
\end{equation}
should not spoil the CMB anisotropy spectrum\cite{Lattanzi:2007ux,Lattanzi:2014mia}. This gives constraints on the mass $m_{\eta}$ and the symmetry breaking scale $f$, as showed in Fig.\ref{fig:majorondm}

As in the axion case there are additional constraints from supernova cooling \cite{Choi:1989hi}, but they are much weaker and give an upper bound much lower than the range shown in Fig.~\ref{fig:majorondm}. 
In the unconstrained region in Fig.~\ref{fig:majorondm}, ARS leptogenesis and majoron DM could in principle work provided the mechanism to generate the majoron mass does not involve further interactions of the sterile neutrinos. 

\subsubsection{Sterile Neutrino}

 We want now to consider the sterile neutrino as a dark matter candidate, a possibility already explored in \cite{Kusenko:2006rh,Petraki:2007gq} in a model with a real scalar field and therefore no Majoron. In our case the presence of the Majoron could make the sterile neutrino unstable, given that it would decay through the channel
\begin{equation}
\Gamma(N\rightarrow\nu\eta)=\frac{1}{32\pi}\left(\frac{m_N}{f}\right)^2m_{\nu}
\end{equation}
Thus one has to assume that the Majoron has a larger mass so that this decay is  kinematically forbidden. 

As in the $B-L$ case, BO leptogenesis is driven by the other two heavier neutrino states $N_{2,3}$, while $N_{1}$ can be produced through freeze-in from $\sigma\rightarrow N_{1} N_{1}$ decay. 
Assuming $\sigma$ is in thermal equilibrium with the bath the Boltzmann equation describing the evolution of the $N_{1}$ density:
\begin{equation}
\dot{n}_{1}+3Hn_{1}=2\frac{M_{\sigma}^2\Gamma_{{\sigma\rightarrow N_{1}N_{1}}}}{2\pi^2}TK_{1}\left(\frac{M_{\sigma}}{T}\right),
\end{equation}
where we have neglected  Pauli blocking, and the inverse processes. 

Following the standard procedure we end with the contribution to the abundance:
\begin{equation}
\Omega_{N_1}h^2\sim\frac{10^{27}}{g^{3/2}_{*}}\frac{m_{N_1}\Gamma_{{\sigma\rightarrow N_{1}N_{1}}}}{M_\sigma^2}.
\end{equation}
Using
\begin{equation}
\Gamma_{{\sigma\rightarrow N_{1}N_{1}}}=\frac{h_N^2M_\sigma}{16\pi}\left(1-\frac{4m_{N_1}^2}{M_\sigma^2}\right)\sim\frac{h_N^2M_\sigma}{16\pi},
\end{equation}
and requiring that $\Omega_{N_1}h^2$ matches the observed DM we find
\begin{equation}
h_N\sim4.3\cdot10^{-13} \sqrt{\frac{m_{N_1}}{M_\sigma}}\left(\frac{g_{*}(m_{N_1})}{10}\right)^{3/4}.
\end{equation}

\begin{figure}
	\centering
	\includegraphics[width=7.0cm,keepaspectratio]{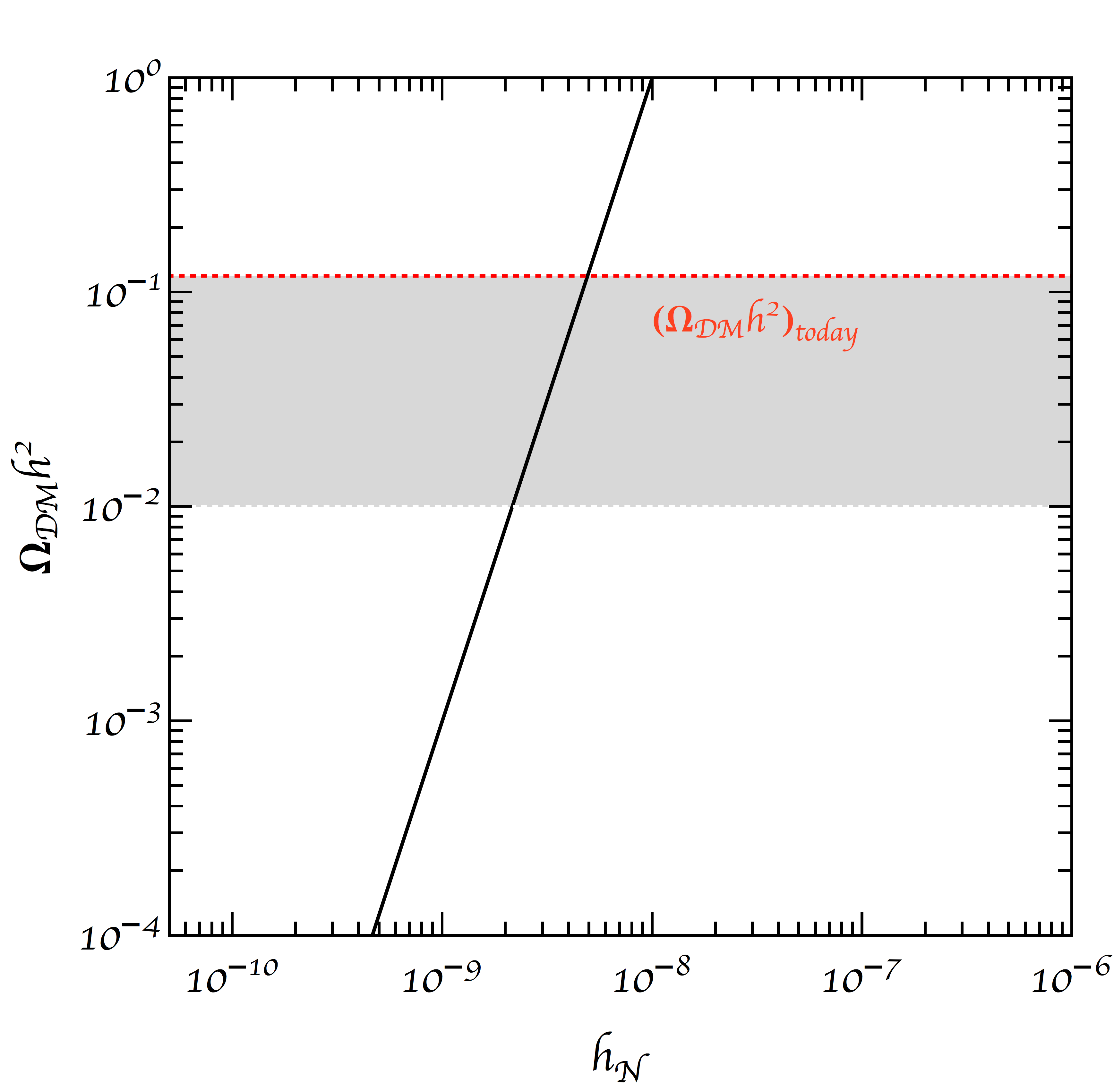}
	\caption{Dark matter relic density as a function of the Yukawa coupling $h_N$ of the lightest sterile neutrino. The gray region identifies $0.01<\Omega_{DM}h^2<0.12$, that is to say a sterile neutrino contribution to DM density between $10$ and $100 \%$. Taking $\langle\Phi\rangle\sim$TeV, this region corresponds to $m_{N_1}\sim keV$ }
	\label{fig:densdark}
\end{figure}

The mass of the DM candidate is related to the coupling which regulates the freeze-in process through the VEV of $\phi$
\begin{eqnarray}
m_{N_1}=h_{N_1} \langle\phi\rangle,
\end{eqnarray}
therefore if $M_\sigma \sim \langle\phi\rangle$, the coupling needs to be
\begin{equation}
h_{N_1}\sim 10^{-9}-10^{-8},
\end{equation}
as shown in  Fig.~\ref{fig:densdark}.

If $m_{N_1}$ is the keV range so that it can satisfy cosmological and astrophysical constraints, the scale of the VEV should be
\begin{equation}
\langle \phi \rangle \sim {\rm TeV}.
\end{equation}
As a consequence we see that if one couples the $B-L$ scalar also to the heavier neutrinos, the interactions are too fast and the BO mechanism cannot work since the bound
in eq.~(\ref{eq:fmaj}) is not satisfied. Alternatively, if we assume eq.~(\ref{eq:fmaj}), then 
\begin{equation}
m_{N}\sim { \mathcal O}({\rm MeV}).
\end{equation}
Such a massive neutrino would need extremely small active-sterile mixing angle (collectively labeled $\theta$) to be sufficiently long-lived. The strongest bound from X-rays \cite{Sekiya:2015jsa} give
\begin{equation}
\sin^2(2\theta)\lesssim {\rm few} \times 10^{-6} \hspace{0.15cm} \hspace{0.15cm}, m_N\sim {\rm keV},
\end{equation}
while the soft gamma ray bound \cite{Boyarsky:2007ge} gives
\begin{equation}
\sin^2(2\theta)\lesssim {\rm few} \times 10^{-21}\hspace{0.15cm}, \hspace{0.15cm} m_N\sim {\rm MeV}.
\end{equation}
In conclusion,  either we consider a  global symmetry with different family charges, more concretely a $(B-L)_1$, or we need to require extremely tiny yukawa coupling for the DM sterile neutrino, making this last model theoretically unappealing.
\section{Conclusion} 
\label{conclusion}

The extension of the Standard Model with three heavy majorana singlets at the weak scale can explain neutrino masses  and also account for the baryon asymmetry in the Universe via the oscillation mechanism \cite{Akhmedov:1998qx,Asaka:2005pn}. This scenario could be testable in future experiments. 
Unfortunately the simplest model cannot easily accommodate dark matter. In the $\nu$MSM \cite{Asaka:2005pn}, one of the three heavy states is in the keV range and provides a candidate for dark matter, but it requires huge lepton asymmetries that cannot be naturally achieved in the minimal setup.
\\In this paper we have explored three extensions of the minimal scenario that 
can accommodate dark matter without spoiling baryogenesis. This is non trivial because new interactions of the heavy singlets can disrupt the necessary out-of-equilibrium condition 
which is mandatory to generate a lepton asymmetry. We have shown that a extension of the minimal model with a $U(1)_{B-L}$ gauge interaction can achieve this goal. The two heavier majorana fermions take part in the generation of the baryon asymmetry, while the lighest one in the keV range, $N_1$, is the dark matter. In contrast with the $\nu$MSM the production 
of the dark matter is not via mixing, but it is dominated by the $B-L$ gauge boson decay. The mixing is however what controls the decay of the $N_1$ and can be made
sufficiently small to avoid the stringent X-ray constraints. The correct DM abundance is achieved for very small $B-L$ gauge couplings, $g_{B-L} \lesssim 10^{-8}$, which are safely small not to disturb the baryon asymmetry, which remains the same as in the minimal model. Such tiny couplings are far below the reach of colliders. Supernova and BBN 
provide the most stringent constraints in the relevant region of parameter space, while future searches in SHIP might have a chance to touch on it.
\\We have also considered an extension involving an invisible axion sector with an extra scalar doublet and  a complex singlet. The heavy majorana singlets get their mass from the PQ breaking scale \cite{Langacker:1986rj}.  DM is in the form of cold axions, from the misalignment mechanism and as is well known, the right relic abundance can be achieved for a large value of the PQ breaking scale, $f_a \simeq 10^{11}$GeV. We have shown that such large scale is compatible with having the heavy neutrinos in the 1-100 GeV scale, and  ARS leptogenesis. Finally we have considered the singlet majoron extension of the minimal model, with a global $U(1)_{B-L}$, that contains two potential DM candidates, the majoron or the lightest   heavy neutrino, $N_1$. Unperturbed ARS baryogenesis requires a relatively high $B-L$ breaking scale, $f \gtrsim 10^{6}$ GeV. Majoron DM requires exotic production scenarios, while neutrino DM works for masses around MeV, which requires  extremely small mixings to make it sufficiently long-lived, or alternatively a less theoretically appealing possibity, where 
the scalar couples to only  one sterile neutrino, while the other two have tree level masses or couple to a different scalar with a larger VEV.\\As a general rule, adding new interactions that affect the heavy Majorana singlets modifies leptogenesis in the minimal model  and viable extensions that can explain DM 
are likely to involve the freeze-in mechanism as in the examples above. 

\section*{Acknowledgements}
	We thank M. Escudero, M. Kekic, J. L\'opez-Pav\'on and J. Salvado  for useful discussions. We acknowledge support from national grants FPA2014-57816-P, FPA2017-85985-P and  the European projects 
	H2020-MSCA-ITN-2015//674896-ELUSIVES and H2020-MSCA-RISE-2015	
\appendix
\section{Computation of momentum averaged rates in gauged $B-L$}

In this appendix we give some details on the computation of the momentum averaged rates in eq.~(\ref{eq:gamv}).  The amplitude for $\bar{f} f\rightarrow N N$ for vanishing masses
is given by
\begin{eqnarray}
\sum_{spins} |{\mathcal M}|^2 = 4 g_{B-L}^4 \sum_f Q_f^2 N_c \left[t^2+u^2\over s^2\right] \equiv A \left[t^2+u^2\over s^2\right], \nonumber\\
\end{eqnarray}
while that for $V V \rightarrow NN$,  in the limit $M^2_\sigma \gg  s,t \gg m_N^2,m_V^2$ is 
\begin{eqnarray}
\sum_{spins} |{\mathcal M}|^2 = 4 g_{B-L}^4 {m_N^2 s \over m_V^4}. 
\end{eqnarray}

Defining the Bose-Einstein and Fermi-Dirac distributions
\begin{eqnarray}
f_B(x) = {1 \over e^x -1}, f_F(x) = {1 \over e^x +1},
\end{eqnarray}
and the variables
\begin{eqnarray}
q_\pm \equiv {1\over 2}(q_0\pm |{\mathbf q}| ),
\end{eqnarray}
where $q= p_1+p_2$. We express all momenta in units of temperature $T$.

Following the procedure of ref.~\cite{Besak:2012qm} the rate $f \bar{f} \rightarrow NN$ can be writen as
\begin{eqnarray}
R^{(1)}(k) = {A \over 4 (2 \pi)^3 k_0} \left(r_1(k)+r_2(k) + r_3(k)\right),
\end{eqnarray}
with
\begin{eqnarray}
r_1(k) \equiv \int_{k_0}^\infty d q_+ \int_0^{k_0} d q_- f_B(q_+ + q_-)  I_1(q_+,q_-),
\end{eqnarray}
\begin{equation}
r_2(k)\equiv 2 \int_{k_0}^\infty d q_+\int_0^{k_0} d q_-f_B(q_+ + q_-)\sum_{i=1,2} I_i(q_+,q_-) a_i[q_+,q_-,k_0] 
\end{equation}
\begin{equation}
r_3(k)\equiv  2\int_{k_0}^\infty d q_+ \int_0^{k_0} d q_-f_B(q_+ + q_-)\sum_{i=1,3} I_i(q_+,q_-) b_i[q_+,q_-,k_0] 
\end{equation}
with
\begin{eqnarray}
I_n(q_+,q_-) \equiv \int_{q_-}^{q_+} x^{n-1} \left[1- 2 f_F(x)\right] d x, 
\end{eqnarray}
and
\begin{eqnarray}
a_1[q_+,q_-,k_0]&\equiv & -1+ {q_+ (k_0-q_-) + q_- (k_0-q_+)\over (q_+-q_-)^2},\nonumber\\
a_2[q_+,q_-,k_0]&\equiv &{q_++q_- -2 k_0 \over (q_+-q_-)^2},\nonumber\\
b_1[q_+,q_-,k_0] &\equiv & a_1^2 + 2 q_+ q_- {(q_+ - k_0) (q_- - k_0) \over (q_+ - q_-)^4}
\nonumber
\end{eqnarray}
\begin{multline}
b_2[q_+,q_-,k_0] \equiv 2 a_1[q_+,q_-,k_0] a_2[q_+,q_-,k_0]\nonumber +\\-2 (q_++ q_-) {(q_+ - k_0) (q_- - k_0) \over (q_+ - q_-)^4}\nonumber\end{multline}
\begin{eqnarray}
b_3[q_+,q_-,k_0] \equiv &  a_2^2+ 2  {(q_+ - k_0) (q_- - k_0) \over (q_+ - q_-)^4}
\end{eqnarray}
The rate $V V \rightarrow NN$ can be written as 
\begin{multline}
R^{(2)}(k)={A'\over(2 \pi)^3k_0} \int_{k_0}^\infty dq_+ \int_0^{k_0} d q_-  (q_+ q_-) f_B(q_+ + q_-)\times \\\ \times I'_1(q_+,q_-),
\end{multline}
with
\begin{eqnarray}
I'_1(q_+,q_-) \equiv \int_{q_-}^{q_+} \left[1+2  f_B(x)\right] d x.
\end{eqnarray}
and $A' \equiv g_{B-L}^4 {m_N^2 \over m_V^4}$.

The total rate is
\begin{eqnarray}
R(k) = \sum_{i=1,2} R^{(i)}(k),
\end{eqnarray}
and the averaged rates $\gamma^{(0)}_V$ and $\gamma^{(1)}_V$ are found to be:
\begin{equation}	
 \langle \gamma^{(0)}_V\rangle =   g_{B-L}^4 T\left(3.2(3) \times 10^{-3}+ 2.95 \times 10^{-4} {m_N^2 T^2\over m_V^4 }\right) 
\end{equation}
\begin{equation}
 \langle \gamma^{(1)}_V\rangle = g_{B-L}^4 T\left(3.4(1) \times 10^{-4} +3.55 \times 10^{-5}  {m_N^2 T^2\over m_V^4 }\right).\end{equation}
where
\begin{equation}
\langle \gamma^{(0)}_V\rangle \equiv  { \int {d^3 k \over (2 \pi)^3 2 k_0} R[k] \over \int {d^3 k \over (2 \pi)^3} f_F(k) }
\end{equation}
and
\begin{equation}
	\langle \gamma^{(1)}_V\rangle \equiv  { \int {d^3 k \over (2 \pi)^3 2 k_0} R[k] f_F(k) \over \int {d^3 k \over (2 \pi)^3} f_F(k)}
\end{equation}

\bibliography{Arsandbeyond.bib}

\begin{thebibliography}{120}
\expandafter\ifx\csname natexlab\endcsname\relax\def\natexlab#1{#1}\fi
\expandafter\ifx\csname bibnamefont\endcsname\relax
  \def\bibnamefont#1{#1}\fi
\expandafter\ifx\csname bibfnamefont\endcsname\relax
  \def\bibfnamefont#1{#1}\fi
\expandafter\ifx\csname citenamefont\endcsname\relax
  \def\citenamefont#1{#1}\fi
\expandafter\ifx\csname url\endcsname\relax
  \def\url#1{\texttt{#1}}\fi
\expandafter\ifx\csname urlprefix\endcsname\relax\def\urlprefix{URL }\fi
\providecommand{\bibinfo}[2]{#2}
\providecommand{\eprint}[2][]{\url{#2}}

\bibitem[{\citenamefont{Minkowski}(1977)}]{Minkowski:1977sc}
\bibinfo{author}{\bibfnamefont{P.}~\bibnamefont{Minkowski}},
  \bibinfo{journal}{Phys. Lett.} \textbf{\bibinfo{volume}{67B}},
  \bibinfo{pages}{421} (\bibinfo{year}{1977}).

\bibitem[{\citenamefont{Gell-Mann et~al.}(1979)\citenamefont{Gell-Mann, Ramond,
  and Slansky}}]{GellMann:1980vs}
\bibinfo{author}{\bibfnamefont{M.}~\bibnamefont{Gell-Mann}},
  \bibinfo{author}{\bibfnamefont{P.}~\bibnamefont{Ramond}}, \bibnamefont{and}
  \bibinfo{author}{\bibfnamefont{R.}~\bibnamefont{Slansky}},
  \bibinfo{journal}{Conf. Proc.} \textbf{\bibinfo{volume}{C790927}},
  \bibinfo{pages}{315} (\bibinfo{year}{1979}), \eprint{1306.4669}.

\bibitem[{\citenamefont{Yanagida}(1979)}]{Yanagida:1979as}
\bibinfo{author}{\bibfnamefont{T.}~\bibnamefont{Yanagida}},
  \bibinfo{journal}{Conf. Proc.} \textbf{\bibinfo{volume}{C7902131}},
  \bibinfo{pages}{95} (\bibinfo{year}{1979}).

\bibitem[{\citenamefont{Mohapatra and Senjanovic}(1980)}]{Mohapatra:1979ia}
\bibinfo{author}{\bibfnamefont{R.~N.} \bibnamefont{Mohapatra}}
  \bibnamefont{and}
  \bibinfo{author}{\bibfnamefont{G.}~\bibnamefont{Senjanovic}},
  \bibinfo{journal}{Phys. Rev. Lett.} \textbf{\bibinfo{volume}{44}},
  \bibinfo{pages}{912} (\bibinfo{year}{1980}).

\bibitem[{\citenamefont{Fukugita and Yanagida}(1986)}]{Fukugita:1986hr}
\bibinfo{author}{\bibfnamefont{M.}~\bibnamefont{Fukugita}} \bibnamefont{and}
  \bibinfo{author}{\bibfnamefont{T.}~\bibnamefont{Yanagida}},
  \bibinfo{journal}{Phys. Lett.} \textbf{\bibinfo{volume}{B174}},
  \bibinfo{pages}{45} (\bibinfo{year}{1986}).

\bibitem[{\citenamefont{Davidson and Ibarra}(2002)}]{Davidson:2002qv}
\bibinfo{author}{\bibfnamefont{S.}~\bibnamefont{Davidson}} \bibnamefont{and}
  \bibinfo{author}{\bibfnamefont{A.}~\bibnamefont{Ibarra}},
  \bibinfo{journal}{Phys. Lett.} \textbf{\bibinfo{volume}{B535}},
  \bibinfo{pages}{25} (\bibinfo{year}{2002}), \eprint{hep-ph/0202239}.

\bibitem[{\citenamefont{Abada et~al.}(2006)\citenamefont{Abada, Davidson,
  Ibarra, Josse-Michaux, Losada, and Riotto}}]{Abada:2006ea}
\bibinfo{author}{\bibfnamefont{A.}~\bibnamefont{Abada}},
  \bibinfo{author}{\bibfnamefont{S.}~\bibnamefont{Davidson}},
  \bibinfo{author}{\bibfnamefont{A.}~\bibnamefont{Ibarra}},
  \bibinfo{author}{\bibfnamefont{F.~X.} \bibnamefont{Josse-Michaux}},
  \bibinfo{author}{\bibfnamefont{M.}~\bibnamefont{Losada}}, \bibnamefont{and}
  \bibinfo{author}{\bibfnamefont{A.}~\bibnamefont{Riotto}},
  \bibinfo{journal}{JHEP} \textbf{\bibinfo{volume}{09}}, \bibinfo{pages}{010}
  (\bibinfo{year}{2006}), \eprint{hep-ph/0605281}.

\bibitem[{\citenamefont{Akhmedov et~al.}(1998)\citenamefont{Akhmedov, Rubakov,
  and Smirnov}}]{Akhmedov:1998qx}
\bibinfo{author}{\bibfnamefont{E.~K.} \bibnamefont{Akhmedov}},
  \bibinfo{author}{\bibfnamefont{V.~A.} \bibnamefont{Rubakov}},
  \bibnamefont{and} \bibinfo{author}{\bibfnamefont{A.~{\relax Yu}.}
  \bibnamefont{Smirnov}}, \bibinfo{journal}{Phys. Rev. Lett.}
  \textbf{\bibinfo{volume}{81}}, \bibinfo{pages}{1359} (\bibinfo{year}{1998}),
  \eprint{hep-ph/9803255}.

\bibitem[{\citenamefont{Asaka and Shaposhnikov}(2005)}]{Asaka:2005pn}
\bibinfo{author}{\bibfnamefont{T.}~\bibnamefont{Asaka}} \bibnamefont{and}
  \bibinfo{author}{\bibfnamefont{M.}~\bibnamefont{Shaposhnikov}},
  \bibinfo{journal}{Phys. Lett.} \textbf{\bibinfo{volume}{B620}},
  \bibinfo{pages}{17} (\bibinfo{year}{2005}), \eprint{hep-ph/0505013}.

\bibitem[{\citenamefont{Kuzmin et~al.}(1985)\citenamefont{Kuzmin, Rubakov, and
  Shaposhnikov}}]{Kuzmin:1985mm}
\bibinfo{author}{\bibfnamefont{V.~A.} \bibnamefont{Kuzmin}},
  \bibinfo{author}{\bibfnamefont{V.~A.} \bibnamefont{Rubakov}},
  \bibnamefont{and} \bibinfo{author}{\bibfnamefont{M.~E.}
  \bibnamefont{Shaposhnikov}}, \bibinfo{journal}{Phys. Lett.}
  \textbf{\bibinfo{volume}{155B}}, \bibinfo{pages}{36} (\bibinfo{year}{1985}).

\bibitem[{\citenamefont{Ferrari et~al.}(2000)\citenamefont{Ferrari, Collot,
  Andrieux, Belhorma, de~Saintignon, Hostachy, Martin, and
  Wielers}}]{Ferrari:2000sp}
\bibinfo{author}{\bibfnamefont{A.}~\bibnamefont{Ferrari}},
  \bibinfo{author}{\bibfnamefont{J.}~\bibnamefont{Collot}},
  \bibinfo{author}{\bibfnamefont{M.-L.} \bibnamefont{Andrieux}},
  \bibinfo{author}{\bibfnamefont{B.}~\bibnamefont{Belhorma}},
  \bibinfo{author}{\bibfnamefont{P.}~\bibnamefont{de~Saintignon}},
  \bibinfo{author}{\bibfnamefont{J.-Y.} \bibnamefont{Hostachy}},
  \bibinfo{author}{\bibfnamefont{P.}~\bibnamefont{Martin}}, \bibnamefont{and}
  \bibinfo{author}{\bibfnamefont{M.}~\bibnamefont{Wielers}},
  \bibinfo{journal}{Phys. Rev.} \textbf{\bibinfo{volume}{D62}},
  \bibinfo{pages}{013001} (\bibinfo{year}{2000}).

\bibitem[{\citenamefont{Graesser}(2007)}]{Graesser:2007pc}
\bibinfo{author}{\bibfnamefont{M.~L.} \bibnamefont{Graesser}}
  (\bibinfo{year}{2007}), \eprint{0705.2190}.

\bibitem[{\citenamefont{del Aguila and
  Aguilar-Saavedra}(2009)}]{delAguila:2008cj}
\bibinfo{author}{\bibfnamefont{F.}~\bibnamefont{del Aguila}} \bibnamefont{and}
  \bibinfo{author}{\bibfnamefont{J.~A.} \bibnamefont{Aguilar-Saavedra}},
  \bibinfo{journal}{Nucl. Phys.} \textbf{\bibinfo{volume}{B813}},
  \bibinfo{pages}{22} (\bibinfo{year}{2009}), \eprint{0808.2468}.

\bibitem[{\citenamefont{Bhupal~Dev et~al.}(2012)\citenamefont{Bhupal~Dev,
  Franceschini, and Mohapatra}}]{BhupalDev:2012zg}
\bibinfo{author}{\bibfnamefont{P.~S.} \bibnamefont{Bhupal~Dev}},
  \bibinfo{author}{\bibfnamefont{R.}~\bibnamefont{Franceschini}},
  \bibnamefont{and} \bibinfo{author}{\bibfnamefont{R.~N.}
  \bibnamefont{Mohapatra}}, \bibinfo{journal}{Phys. Rev.}
  \textbf{\bibinfo{volume}{D86}}, \bibinfo{pages}{093010}
  (\bibinfo{year}{2012}), \eprint{1207.2756}.

\bibitem[{\citenamefont{Helo et~al.}(2014)\citenamefont{Helo, Hirsch, and
  Kovalenko}}]{Helo:2013esa}
\bibinfo{author}{\bibfnamefont{J.~C.} \bibnamefont{Helo}},
  \bibinfo{author}{\bibfnamefont{M.}~\bibnamefont{Hirsch}}, \bibnamefont{and}
  \bibinfo{author}{\bibfnamefont{S.}~\bibnamefont{Kovalenko}},
  \bibinfo{journal}{Phys. Rev.} \textbf{\bibinfo{volume}{D89}},
  \bibinfo{pages}{073005} (\bibinfo{year}{2014}), \bibinfo{note}{[Erratum:
  Phys. Rev.D93,no.9,099902(2016)]}, \eprint{1312.2900}.

\bibitem[{\citenamefont{Blondel et~al.}(2016)\citenamefont{Blondel, Graverini,
  Serra, and Shaposhnikov}}]{Blondel:2014bra}
\bibinfo{author}{\bibfnamefont{A.}~\bibnamefont{Blondel}},
  \bibinfo{author}{\bibfnamefont{E.}~\bibnamefont{Graverini}},
  \bibinfo{author}{\bibfnamefont{N.}~\bibnamefont{Serra}}, \bibnamefont{and}
  \bibinfo{author}{\bibfnamefont{M.}~\bibnamefont{Shaposhnikov}}
  (\bibinfo{collaboration}{FCC-ee study Team}), \bibinfo{journal}{Nucl. Part.
  Phys. Proc.} \textbf{\bibinfo{volume}{273-275}}, \bibinfo{pages}{1883}
  (\bibinfo{year}{2016}), \eprint{1411.5230}.

\bibitem[{\citenamefont{Abada et~al.}(2015{\natexlab{a}})\citenamefont{Abada,
  De~Romeri, Monteil, Orloff, and Teixeira}}]{Abada:2014cca}
\bibinfo{author}{\bibfnamefont{A.}~\bibnamefont{Abada}},
  \bibinfo{author}{\bibfnamefont{V.}~\bibnamefont{De~Romeri}},
  \bibinfo{author}{\bibfnamefont{S.}~\bibnamefont{Monteil}},
  \bibinfo{author}{\bibfnamefont{J.}~\bibnamefont{Orloff}}, \bibnamefont{and}
  \bibinfo{author}{\bibfnamefont{A.~M.} \bibnamefont{Teixeira}},
  \bibinfo{journal}{JHEP} \textbf{\bibinfo{volume}{04}}, \bibinfo{pages}{051}
  (\bibinfo{year}{2015}{\natexlab{a}}), \eprint{1412.6322}.

\bibitem[{\citenamefont{Cui and Shuve}(2015)}]{Cui:2014twa}
\bibinfo{author}{\bibfnamefont{Y.}~\bibnamefont{Cui}} \bibnamefont{and}
  \bibinfo{author}{\bibfnamefont{B.}~\bibnamefont{Shuve}},
  \bibinfo{journal}{JHEP} \textbf{\bibinfo{volume}{02}}, \bibinfo{pages}{049}
  (\bibinfo{year}{2015}), \eprint{1409.6729}.

\bibitem[{\citenamefont{Antusch and Fischer}(2015)}]{Antusch:2015mia}
\bibinfo{author}{\bibfnamefont{S.}~\bibnamefont{Antusch}} \bibnamefont{and}
  \bibinfo{author}{\bibfnamefont{O.}~\bibnamefont{Fischer}},
  \bibinfo{journal}{JHEP} \textbf{\bibinfo{volume}{05}}, \bibinfo{pages}{053}
  (\bibinfo{year}{2015}), \eprint{1502.05915}.

\bibitem[{\citenamefont{Gago et~al.}(2015)\citenamefont{Gago, Hernandez,
  Jones-Perez, Losada, and Moreno~Briceño}}]{Gago:2015vma}
\bibinfo{author}{\bibfnamefont{A.~M.} \bibnamefont{Gago}},
  \bibinfo{author}{\bibfnamefont{P.}~\bibnamefont{Hernandez}},
  \bibinfo{author}{\bibfnamefont{J.}~\bibnamefont{Jones-Perez}},
  \bibinfo{author}{\bibfnamefont{M.}~\bibnamefont{Losada}}, \bibnamefont{and}
  \bibinfo{author}{\bibfnamefont{A.}~\bibnamefont{Moreno~Briceño}},
  \bibinfo{journal}{Eur. Phys. J.} \textbf{\bibinfo{volume}{C75}},
  \bibinfo{pages}{470} (\bibinfo{year}{2015}), \eprint{1505.05880}.

\bibitem[{\citenamefont{Antusch et~al.}(2016)\citenamefont{Antusch, Cazzato,
  and Fischer}}]{Antusch:2016vyf}
\bibinfo{author}{\bibfnamefont{S.}~\bibnamefont{Antusch}},
  \bibinfo{author}{\bibfnamefont{E.}~\bibnamefont{Cazzato}}, \bibnamefont{and}
  \bibinfo{author}{\bibfnamefont{O.}~\bibnamefont{Fischer}},
  \bibinfo{journal}{JHEP} \textbf{\bibinfo{volume}{12}}, \bibinfo{pages}{007}
  (\bibinfo{year}{2016}), \eprint{1604.02420}.

\bibitem[{\citenamefont{Caputo et~al.}(2017{\natexlab{a}})\citenamefont{Caputo,
  Hernandez, Kekic, Lopez-Pavon, and Salvado}}]{Caputo:2016ojx}
\bibinfo{author}{\bibfnamefont{A.}~\bibnamefont{Caputo}},
  \bibinfo{author}{\bibfnamefont{P.}~\bibnamefont{Hernandez}},
  \bibinfo{author}{\bibfnamefont{M.}~\bibnamefont{Kekic}},
  \bibinfo{author}{\bibfnamefont{J.}~\bibnamefont{Lopez-Pavon}},
  \bibnamefont{and} \bibinfo{author}{\bibfnamefont{J.}~\bibnamefont{Salvado}},
  \bibinfo{journal}{Eur. Phys. J.} \textbf{\bibinfo{volume}{C77}},
  \bibinfo{pages}{258} (\bibinfo{year}{2017}{\natexlab{a}}),
  \eprint{1611.05000}.

\bibitem[{\citenamefont{Caputo et~al.}(2017{\natexlab{b}})\citenamefont{Caputo,
  Hernandez, Lopez-Pavon, and Salvado}}]{Caputo:2017pit}
\bibinfo{author}{\bibfnamefont{A.}~\bibnamefont{Caputo}},
  \bibinfo{author}{\bibfnamefont{P.}~\bibnamefont{Hernandez}},
  \bibinfo{author}{\bibfnamefont{J.}~\bibnamefont{Lopez-Pavon}},
  \bibnamefont{and} \bibinfo{author}{\bibfnamefont{J.}~\bibnamefont{Salvado}},
  \bibinfo{journal}{JHEP} \textbf{\bibinfo{volume}{06}}, \bibinfo{pages}{112}
  (\bibinfo{year}{2017}{\natexlab{b}}), \eprint{1704.08721}.

\bibitem[{\citenamefont{Shaposhnikov}(2008)}]{Shaposhnikov:2008pf}
\bibinfo{author}{\bibfnamefont{M.}~\bibnamefont{Shaposhnikov}},
  \bibinfo{journal}{JHEP} \textbf{\bibinfo{volume}{08}}, \bibinfo{pages}{008}
  (\bibinfo{year}{2008}), \eprint{0804.4542}.

\bibitem[{\citenamefont{Canetti et~al.}(2012)\citenamefont{Canetti, Drewes, and
  Shaposhnikov}}]{Canetti:2012zc}
\bibinfo{author}{\bibfnamefont{L.}~\bibnamefont{Canetti}},
  \bibinfo{author}{\bibfnamefont{M.}~\bibnamefont{Drewes}}, \bibnamefont{and}
  \bibinfo{author}{\bibfnamefont{M.}~\bibnamefont{Shaposhnikov}},
  \bibinfo{journal}{New J. Phys.} \textbf{\bibinfo{volume}{14}},
  \bibinfo{pages}{095012} (\bibinfo{year}{2012}), \eprint{1204.4186}.

\bibitem[{\citenamefont{Canetti et~al.}(2013)\citenamefont{Canetti, Drewes,
  Frossard, and Shaposhnikov}}]{Canetti:2012kh}
\bibinfo{author}{\bibfnamefont{L.}~\bibnamefont{Canetti}},
  \bibinfo{author}{\bibfnamefont{M.}~\bibnamefont{Drewes}},
  \bibinfo{author}{\bibfnamefont{T.}~\bibnamefont{Frossard}}, \bibnamefont{and}
  \bibinfo{author}{\bibfnamefont{M.}~\bibnamefont{Shaposhnikov}},
  \bibinfo{journal}{Phys. Rev.} \textbf{\bibinfo{volume}{D87}},
  \bibinfo{pages}{093006} (\bibinfo{year}{2013}), \eprint{1208.4607}.

\bibitem[{\citenamefont{Asaka et~al.}(2012)\citenamefont{Asaka, Eijima, and
  Ishida}}]{Asaka:2011wq}
\bibinfo{author}{\bibfnamefont{T.}~\bibnamefont{Asaka}},
  \bibinfo{author}{\bibfnamefont{S.}~\bibnamefont{Eijima}}, \bibnamefont{and}
  \bibinfo{author}{\bibfnamefont{H.}~\bibnamefont{Ishida}},
  \bibinfo{journal}{JCAP} \textbf{\bibinfo{volume}{1202}}, \bibinfo{pages}{021}
  (\bibinfo{year}{2012}), \eprint{1112.5565}.

\bibitem[{\citenamefont{Shuve and Yavin}(2014)}]{Shuve:2014zua}
\bibinfo{author}{\bibfnamefont{B.}~\bibnamefont{Shuve}} \bibnamefont{and}
  \bibinfo{author}{\bibfnamefont{I.}~\bibnamefont{Yavin}},
  \bibinfo{journal}{Phys. Rev.} \textbf{\bibinfo{volume}{D89}},
  \bibinfo{pages}{075014} (\bibinfo{year}{2014}), \eprint{1401.2459}.

\bibitem[{\citenamefont{Abada et~al.}(2015{\natexlab{b}})\citenamefont{Abada,
  Arcadi, Domcke, and Lucente}}]{Abada:2015rta}
\bibinfo{author}{\bibfnamefont{A.}~\bibnamefont{Abada}},
  \bibinfo{author}{\bibfnamefont{G.}~\bibnamefont{Arcadi}},
  \bibinfo{author}{\bibfnamefont{V.}~\bibnamefont{Domcke}}, \bibnamefont{and}
  \bibinfo{author}{\bibfnamefont{M.}~\bibnamefont{Lucente}},
  \bibinfo{journal}{JCAP} \textbf{\bibinfo{volume}{1511}}, \bibinfo{pages}{041}
  (\bibinfo{year}{2015}{\natexlab{b}}), \eprint{1507.06215}.

\bibitem[{\citenamefont{Hernandez et~al.}(2015)\citenamefont{Hernandez, Kekic,
  Lopez-Pavon, Racker, and Rius}}]{Hernandez:2015wna}
\bibinfo{author}{\bibfnamefont{P.}~\bibnamefont{Hernandez}},
  \bibinfo{author}{\bibfnamefont{M.}~\bibnamefont{Kekic}},
  \bibinfo{author}{\bibfnamefont{J.}~\bibnamefont{Lopez-Pavon}},
  \bibinfo{author}{\bibfnamefont{J.}~\bibnamefont{Racker}}, \bibnamefont{and}
  \bibinfo{author}{\bibfnamefont{N.}~\bibnamefont{Rius}},
  \bibinfo{journal}{JHEP} \textbf{\bibinfo{volume}{10}}, \bibinfo{pages}{067}
  (\bibinfo{year}{2015}), \eprint{1508.03676}.

\bibitem[{\citenamefont{Hernandez et~al.}(2016)\citenamefont{Hernandez, Kekic,
  Lopez-Pavon, Racker, and Salvado}}]{Hernandez:2016kel}
\bibinfo{author}{\bibfnamefont{P.}~\bibnamefont{Hernandez}},
  \bibinfo{author}{\bibfnamefont{M.}~\bibnamefont{Kekic}},
  \bibinfo{author}{\bibfnamefont{J.}~\bibnamefont{Lopez-Pavon}},
  \bibinfo{author}{\bibfnamefont{J.}~\bibnamefont{Racker}}, \bibnamefont{and}
  \bibinfo{author}{\bibfnamefont{J.}~\bibnamefont{Salvado}},
  \bibinfo{journal}{JHEP} \textbf{\bibinfo{volume}{08}}, \bibinfo{pages}{157}
  (\bibinfo{year}{2016}), \eprint{1606.06719}.

\bibitem[{\citenamefont{Drewes et~al.}(2016)\citenamefont{Drewes, Garbrecht,
  Gueter, and Klaric}}]{Drewes:2016gmt}
\bibinfo{author}{\bibfnamefont{M.}~\bibnamefont{Drewes}},
  \bibinfo{author}{\bibfnamefont{B.}~\bibnamefont{Garbrecht}},
  \bibinfo{author}{\bibfnamefont{D.}~\bibnamefont{Gueter}}, \bibnamefont{and}
  \bibinfo{author}{\bibfnamefont{J.}~\bibnamefont{Klaric}},
  \bibinfo{journal}{JHEP} \textbf{\bibinfo{volume}{12}}, \bibinfo{pages}{150}
  (\bibinfo{year}{2016}), \eprint{1606.06690}.

\bibitem[{\citenamefont{Drewes et~al.}(2017)\citenamefont{Drewes, Garbrecht,
  Gueter, and Klaric}}]{Drewes:2016jae}
\bibinfo{author}{\bibfnamefont{M.}~\bibnamefont{Drewes}},
  \bibinfo{author}{\bibfnamefont{B.}~\bibnamefont{Garbrecht}},
  \bibinfo{author}{\bibfnamefont{D.}~\bibnamefont{Gueter}}, \bibnamefont{and}
  \bibinfo{author}{\bibfnamefont{J.}~\bibnamefont{Klaric}},
  \bibinfo{journal}{JHEP} \textbf{\bibinfo{volume}{08}}, \bibinfo{pages}{018}
  (\bibinfo{year}{2017}), \eprint{1609.09069}.

\bibitem[{\citenamefont{Hambye and Teresi}(2016)}]{Hambye:2016sby}
\bibinfo{author}{\bibfnamefont{T.}~\bibnamefont{Hambye}} \bibnamefont{and}
  \bibinfo{author}{\bibfnamefont{D.}~\bibnamefont{Teresi}},
  \bibinfo{journal}{Phys. Rev. Lett.} \textbf{\bibinfo{volume}{117}},
  \bibinfo{pages}{091801} (\bibinfo{year}{2016}), \eprint{1606.00017}.

\bibitem[{\citenamefont{Ghiglieri and Laine}(2017)}]{Ghiglieri:2017gjz}
\bibinfo{author}{\bibfnamefont{J.}~\bibnamefont{Ghiglieri}} \bibnamefont{and}
  \bibinfo{author}{\bibfnamefont{M.}~\bibnamefont{Laine}},
  \bibinfo{journal}{JHEP} \textbf{\bibinfo{volume}{05}}, \bibinfo{pages}{132}
  (\bibinfo{year}{2017}), \eprint{1703.06087}.

\bibitem[{\citenamefont{Asaka et~al.}(2017)\citenamefont{Asaka, Eijima, Ishida,
  Minogawa, and Yoshii}}]{Asaka:2017rdj}
\bibinfo{author}{\bibfnamefont{T.}~\bibnamefont{Asaka}},
  \bibinfo{author}{\bibfnamefont{S.}~\bibnamefont{Eijima}},
  \bibinfo{author}{\bibfnamefont{H.}~\bibnamefont{Ishida}},
  \bibinfo{author}{\bibfnamefont{K.}~\bibnamefont{Minogawa}}, \bibnamefont{and}
  \bibinfo{author}{\bibfnamefont{T.}~\bibnamefont{Yoshii}}
  (\bibinfo{year}{2017}), \eprint{1704.02692}.

\bibitem[{\citenamefont{Hambye and Teresi}(2017)}]{Hambye:2017elz}
\bibinfo{author}{\bibfnamefont{T.}~\bibnamefont{Hambye}} \bibnamefont{and}
  \bibinfo{author}{\bibfnamefont{D.}~\bibnamefont{Teresi}},
  \bibinfo{journal}{Phys. Rev.} \textbf{\bibinfo{volume}{D96}},
  \bibinfo{pages}{015031} (\bibinfo{year}{2017}), \eprint{1705.00016}.

\bibitem[{\citenamefont{Abada et~al.}(2017)\citenamefont{Abada, Arcadi, Domcke,
  and Lucente}}]{Abada:2017ieq}
\bibinfo{author}{\bibfnamefont{A.}~\bibnamefont{Abada}},
  \bibinfo{author}{\bibfnamefont{G.}~\bibnamefont{Arcadi}},
  \bibinfo{author}{\bibfnamefont{V.}~\bibnamefont{Domcke}}, \bibnamefont{and}
  \bibinfo{author}{\bibfnamefont{M.}~\bibnamefont{Lucente}},
  \bibinfo{journal}{JCAP} \textbf{\bibinfo{volume}{1712}}, \bibinfo{pages}{024}
  (\bibinfo{year}{2017}), \eprint{1709.00415}.

\bibitem[{\citenamefont{Ghiglieri and Laine}(2018)}]{Ghiglieri:2017csp}
\bibinfo{author}{\bibfnamefont{J.}~\bibnamefont{Ghiglieri}} \bibnamefont{and}
  \bibinfo{author}{\bibfnamefont{M.}~\bibnamefont{Laine}},
  \bibinfo{journal}{JHEP} \textbf{\bibinfo{volume}{02}}, \bibinfo{pages}{078}
  (\bibinfo{year}{2018}), \eprint{1711.08469}.

\bibitem[{\citenamefont{Dodelson and Widrow}(1994)}]{Dodelson:1993je}
\bibinfo{author}{\bibfnamefont{S.}~\bibnamefont{Dodelson}} \bibnamefont{and}
  \bibinfo{author}{\bibfnamefont{L.~M.} \bibnamefont{Widrow}},
  \bibinfo{journal}{Phys. Rev. Lett.} \textbf{\bibinfo{volume}{72}},
  \bibinfo{pages}{17} (\bibinfo{year}{1994}), \eprint{hep-ph/9303287}.

\bibitem[{\citenamefont{Shi and Fuller}(1999)}]{Shi:1998km}
\bibinfo{author}{\bibfnamefont{X.-D.} \bibnamefont{Shi}} \bibnamefont{and}
  \bibinfo{author}{\bibfnamefont{G.~M.} \bibnamefont{Fuller}},
  \bibinfo{journal}{Phys. Rev. Lett.} \textbf{\bibinfo{volume}{82}},
  \bibinfo{pages}{2832} (\bibinfo{year}{1999}), \eprint{astro-ph/9810076}.

\bibitem[{\citenamefont{Perez et~al.}(2017)\citenamefont{Perez, Ng, Beacom,
  Hersh, Horiuchi, and Krivonos}}]{Perez:2016tcq}
\bibinfo{author}{\bibfnamefont{K.}~\bibnamefont{Perez}},
  \bibinfo{author}{\bibfnamefont{K.~C.~Y.} \bibnamefont{Ng}},
  \bibinfo{author}{\bibfnamefont{J.~F.} \bibnamefont{Beacom}},
  \bibinfo{author}{\bibfnamefont{C.}~\bibnamefont{Hersh}},
  \bibinfo{author}{\bibfnamefont{S.}~\bibnamefont{Horiuchi}}, \bibnamefont{and}
  \bibinfo{author}{\bibfnamefont{R.}~\bibnamefont{Krivonos}},
  \bibinfo{journal}{Phys. Rev.} \textbf{\bibinfo{volume}{D95}},
  \bibinfo{pages}{123002} (\bibinfo{year}{2017}), \eprint{1609.00667}.

\bibitem[{\citenamefont{Baur et~al.}(2017)\citenamefont{Baur,
  Palanque-Delabrouille, Yeche, Boyarsky, Ruchayskiy, Armengaud, and
  Lesgourgues}}]{Baur:2017stq}
\bibinfo{author}{\bibfnamefont{J.}~\bibnamefont{Baur}},
  \bibinfo{author}{\bibfnamefont{N.}~\bibnamefont{Palanque-Delabrouille}},
  \bibinfo{author}{\bibfnamefont{C.}~\bibnamefont{Yeche}},
  \bibinfo{author}{\bibfnamefont{A.}~\bibnamefont{Boyarsky}},
  \bibinfo{author}{\bibfnamefont{O.}~\bibnamefont{Ruchayskiy}},
  \bibinfo{author}{\bibfnamefont{Ã.}~\bibnamefont{Armengaud}},
  \bibnamefont{and}
  \bibinfo{author}{\bibfnamefont{J.}~\bibnamefont{Lesgourgues}},
  \bibinfo{journal}{JCAP} \textbf{\bibinfo{volume}{1712}}, \bibinfo{pages}{013}
  (\bibinfo{year}{2017}), \eprint{1706.03118}.

\bibitem[{\citenamefont{Mohapatra and Marshak}(1980)}]{Mohapatra:1980qe}
\bibinfo{author}{\bibfnamefont{R.~N.} \bibnamefont{Mohapatra}}
  \bibnamefont{and} \bibinfo{author}{\bibfnamefont{R.~E.}
  \bibnamefont{Marshak}}, \bibinfo{journal}{Phys. Rev. Lett.}
  \textbf{\bibinfo{volume}{44}}, \bibinfo{pages}{1316} (\bibinfo{year}{1980}),
  \bibinfo{note}{[Erratum: Phys. Rev. Lett.44,1643(1980)]}.

\bibitem[{\citenamefont{Langacker et~al.}(1986)\citenamefont{Langacker, Peccei,
  and Yanagida}}]{Langacker:1986rj}
\bibinfo{author}{\bibfnamefont{P.}~\bibnamefont{Langacker}},
  \bibinfo{author}{\bibfnamefont{R.~D.} \bibnamefont{Peccei}},
  \bibnamefont{and} \bibinfo{author}{\bibfnamefont{T.}~\bibnamefont{Yanagida}},
  \bibinfo{journal}{Mod. Phys. Lett.} \textbf{\bibinfo{volume}{A1}},
  \bibinfo{pages}{541} (\bibinfo{year}{1986}).

\bibitem[{\citenamefont{Chikashige et~al.}(1981)\citenamefont{Chikashige,
  Mohapatra, and Peccei}}]{Chikashige:1980ui}
\bibinfo{author}{\bibfnamefont{Y.}~\bibnamefont{Chikashige}},
  \bibinfo{author}{\bibfnamefont{R.~N.} \bibnamefont{Mohapatra}},
  \bibnamefont{and} \bibinfo{author}{\bibfnamefont{R.~D.}
  \bibnamefont{Peccei}}, \bibinfo{journal}{Phys. Lett.}
  \textbf{\bibinfo{volume}{98B}}, \bibinfo{pages}{265} (\bibinfo{year}{1981}).

\bibitem[{\citenamefont{Schechter and Valle}(1982)}]{Schechter:1981cv}
\bibinfo{author}{\bibfnamefont{J.}~\bibnamefont{Schechter}} \bibnamefont{and}
  \bibinfo{author}{\bibfnamefont{J.~W.~F.} \bibnamefont{Valle}},
  \bibinfo{journal}{Phys. Rev.} \textbf{\bibinfo{volume}{D25}},
  \bibinfo{pages}{774} (\bibinfo{year}{1982}).

\bibitem[{\citenamefont{Drewes et~al.}(2018)\citenamefont{Drewes, Garbrecht,
  Hernandez, Kekic, Lopez-Pavon, Racker, Rius, Salvado, and
  Teresi}}]{Drewes:2017zyw}
\bibinfo{author}{\bibfnamefont{M.}~\bibnamefont{Drewes}},
  \bibinfo{author}{\bibfnamefont{B.}~\bibnamefont{Garbrecht}},
  \bibinfo{author}{\bibfnamefont{P.}~\bibnamefont{Hernandez}},
  \bibinfo{author}{\bibfnamefont{M.}~\bibnamefont{Kekic}},
  \bibinfo{author}{\bibfnamefont{J.}~\bibnamefont{Lopez-Pavon}},
  \bibinfo{author}{\bibfnamefont{J.}~\bibnamefont{Racker}},
  \bibinfo{author}{\bibfnamefont{N.}~\bibnamefont{Rius}},
  \bibinfo{author}{\bibfnamefont{J.}~\bibnamefont{Salvado}}, \bibnamefont{and}
  \bibinfo{author}{\bibfnamefont{D.}~\bibnamefont{Teresi}},
  \bibinfo{journal}{Int. J. Mod. Phys.} \textbf{\bibinfo{volume}{A33}},
  \bibinfo{pages}{1842002} (\bibinfo{year}{2018}), \eprint{1711.02862}.

\bibitem[{\citenamefont{Besak and Bodeker}(2012)}]{Besak:2012qm}
\bibinfo{author}{\bibfnamefont{D.}~\bibnamefont{Besak}} \bibnamefont{and}
  \bibinfo{author}{\bibfnamefont{D.}~\bibnamefont{Bodeker}},
  \bibinfo{journal}{JCAP} \textbf{\bibinfo{volume}{1203}}, \bibinfo{pages}{029}
  (\bibinfo{year}{2012}), \eprint{1202.1288}.

\bibitem[{\citenamefont{Garbrecht et~al.}(2013)\citenamefont{Garbrecht, Glowna,
  and Schwaller}}]{Garbrecht:2013urw}
\bibinfo{author}{\bibfnamefont{B.}~\bibnamefont{Garbrecht}},
  \bibinfo{author}{\bibfnamefont{F.}~\bibnamefont{Glowna}}, \bibnamefont{and}
  \bibinfo{author}{\bibfnamefont{P.}~\bibnamefont{Schwaller}},
  \bibinfo{journal}{Nucl. Phys.} \textbf{\bibinfo{volume}{B877}},
  \bibinfo{pages}{1} (\bibinfo{year}{2013}), \eprint{1303.5498}.

\bibitem[{\citenamefont{Ghisoiu and Laine}(2014)}]{Ghisoiu:2014ena}
\bibinfo{author}{\bibfnamefont{I.}~\bibnamefont{Ghisoiu}} \bibnamefont{and}
  \bibinfo{author}{\bibfnamefont{M.}~\bibnamefont{Laine}},
  \bibinfo{journal}{JCAP} \textbf{\bibinfo{volume}{1412}}, \bibinfo{pages}{032}
  (\bibinfo{year}{2014}), \eprint{1411.1765}.

\bibitem[{\citenamefont{Bellini et~al.}(2011)}]{Bellini:2011rx}
\bibinfo{author}{\bibfnamefont{G.}~\bibnamefont{Bellini}} \bibnamefont{et~al.},
  \bibinfo{journal}{Phys. Rev. Lett.} \textbf{\bibinfo{volume}{107}},
  \bibinfo{pages}{141302} (\bibinfo{year}{2011}), \eprint{1104.1816}.

\bibitem[{\citenamefont{Chatrchyan et~al.}(2013)}]{Chatrchyan:2013tia}
\bibinfo{author}{\bibfnamefont{S.}~\bibnamefont{Chatrchyan}}
  \bibnamefont{et~al.} (\bibinfo{collaboration}{CMS}), \bibinfo{journal}{JHEP}
  \textbf{\bibinfo{volume}{12}}, \bibinfo{pages}{030} (\bibinfo{year}{2013}),
  \eprint{1310.7291}.

\bibitem[{\citenamefont{Lees et~al.}(2014)}]{Lees:2014xha}
\bibinfo{author}{\bibfnamefont{J.~P.} \bibnamefont{Lees}} \bibnamefont{et~al.}
  (\bibinfo{collaboration}{BaBar}), \bibinfo{journal}{Phys. Rev. Lett.}
  \textbf{\bibinfo{volume}{113}}, \bibinfo{pages}{201801}
  (\bibinfo{year}{2014}), \eprint{1406.2980}.

\bibitem[{\citenamefont{Lees et~al.}(2017)}]{Lees:2017lec}
\bibinfo{author}{\bibfnamefont{J.~P.} \bibnamefont{Lees}} \bibnamefont{et~al.}
  (\bibinfo{collaboration}{BaBar}), \bibinfo{journal}{Phys. Rev. Lett.}
  \textbf{\bibinfo{volume}{119}}, \bibinfo{pages}{131804}
  (\bibinfo{year}{2017}), \eprint{1702.03327}.

\bibitem[{\citenamefont{Khachatryan et~al.}(2015)}]{Khachatryan:2014fba}
\bibinfo{author}{\bibfnamefont{V.}~\bibnamefont{Khachatryan}}
  \bibnamefont{et~al.} (\bibinfo{collaboration}{CMS}), \bibinfo{journal}{JHEP}
  \textbf{\bibinfo{volume}{04}}, \bibinfo{pages}{025} (\bibinfo{year}{2015}),
  \eprint{1412.6302}.

\bibitem[{\citenamefont{Aad et~al.}(2014)}]{Aad:2014cka}
\bibinfo{author}{\bibfnamefont{G.}~\bibnamefont{Aad}} \bibnamefont{et~al.}
  (\bibinfo{collaboration}{ATLAS}), \bibinfo{journal}{Phys. Rev.}
  \textbf{\bibinfo{volume}{D90}}, \bibinfo{pages}{052005}
  (\bibinfo{year}{2014}), \eprint{1405.4123}.

\bibitem[{ALE(2004)}]{ALEPH:2004aa}
 (\bibinfo{year}{2004}), \eprint{hep-ex/0412015}.

\bibitem[{\citenamefont{Appelquist et~al.}(2003)\citenamefont{Appelquist,
  Dobrescu, and Hopper}}]{Appelquist:2002mw}
\bibinfo{author}{\bibfnamefont{T.}~\bibnamefont{Appelquist}},
  \bibinfo{author}{\bibfnamefont{B.~A.} \bibnamefont{Dobrescu}},
  \bibnamefont{and} \bibinfo{author}{\bibfnamefont{A.~R.}
  \bibnamefont{Hopper}}, \bibinfo{journal}{Phys. Rev.}
  \textbf{\bibinfo{volume}{D68}}, \bibinfo{pages}{035012}
  (\bibinfo{year}{2003}), \eprint{hep-ph/0212073}.

\bibitem[{\citenamefont{Batell et~al.}(2016)\citenamefont{Batell, Pospelov, and
  Shuve}}]{Batell:2016zod}
\bibinfo{author}{\bibfnamefont{B.}~\bibnamefont{Batell}},
  \bibinfo{author}{\bibfnamefont{M.}~\bibnamefont{Pospelov}}, \bibnamefont{and}
  \bibinfo{author}{\bibfnamefont{B.}~\bibnamefont{Shuve}},
  \bibinfo{journal}{JHEP} \textbf{\bibinfo{volume}{08}}, \bibinfo{pages}{052}
  (\bibinfo{year}{2016}), \eprint{1604.06099}.

\bibitem[{\citenamefont{Klasen et~al.}(2017)\citenamefont{Klasen, Lyonnet, and
  Queiroz}}]{Klasen:2016qux}
\bibinfo{author}{\bibfnamefont{M.}~\bibnamefont{Klasen}},
  \bibinfo{author}{\bibfnamefont{F.}~\bibnamefont{Lyonnet}}, \bibnamefont{and}
  \bibinfo{author}{\bibfnamefont{F.~S.} \bibnamefont{Queiroz}},
  \bibinfo{journal}{Eur. Phys. J.} \textbf{\bibinfo{volume}{C77}},
  \bibinfo{pages}{348} (\bibinfo{year}{2017}), \eprint{1607.06468}.

\bibitem[{\citenamefont{Ilten et~al.}(2018)\citenamefont{Ilten, Soreq,
  Williams, and Xue}}]{Ilten:2018crw}
\bibinfo{author}{\bibfnamefont{P.}~\bibnamefont{Ilten}},
  \bibinfo{author}{\bibfnamefont{Y.}~\bibnamefont{Soreq}},
  \bibinfo{author}{\bibfnamefont{M.}~\bibnamefont{Williams}}, \bibnamefont{and}
  \bibinfo{author}{\bibfnamefont{W.}~\bibnamefont{Xue}},
  \bibinfo{journal}{JHEP} \textbf{\bibinfo{volume}{06}}, \bibinfo{pages}{004}
  (\bibinfo{year}{2018}), \eprint{1801.04847}.

\bibitem[{\citenamefont{Escudero et~al.}(2018)\citenamefont{Escudero, Witte,
  and Rius}}]{Escudero:2018fwn}
\bibinfo{author}{\bibfnamefont{M.}~\bibnamefont{Escudero}},
  \bibinfo{author}{\bibfnamefont{S.~J.} \bibnamefont{Witte}}, \bibnamefont{and}
  \bibinfo{author}{\bibfnamefont{N.}~\bibnamefont{Rius}}
  (\bibinfo{year}{2018}), \eprint{1806.02823}.

\bibitem[{\citenamefont{Alekhin et~al.}(2016)}]{Alekhin:2015byh}
\bibinfo{author}{\bibfnamefont{S.}~\bibnamefont{Alekhin}} \bibnamefont{et~al.},
  \bibinfo{journal}{Rept. Prog. Phys.} \textbf{\bibinfo{volume}{79}},
  \bibinfo{pages}{124201} (\bibinfo{year}{2016}), \eprint{1504.04855}.

\bibitem[{\citenamefont{Gorbunov et~al.}(2015)\citenamefont{Gorbunov, Makarov,
  and Timiryasov}}]{Gorbunov:2014wqa}
\bibinfo{author}{\bibfnamefont{D.}~\bibnamefont{Gorbunov}},
  \bibinfo{author}{\bibfnamefont{A.}~\bibnamefont{Makarov}}, \bibnamefont{and}
  \bibinfo{author}{\bibfnamefont{I.}~\bibnamefont{Timiryasov}},
  \bibinfo{journal}{Phys. Rev.} \textbf{\bibinfo{volume}{D91}},
  \bibinfo{pages}{035027} (\bibinfo{year}{2015}), \eprint{1411.4007}.

\bibitem[{\citenamefont{Kaneta et~al.}(2017)\citenamefont{Kaneta, Kang, and
  Lee}}]{Kaneta:2016vkq}
\bibinfo{author}{\bibfnamefont{K.}~\bibnamefont{Kaneta}},
  \bibinfo{author}{\bibfnamefont{Z.}~\bibnamefont{Kang}}, \bibnamefont{and}
  \bibinfo{author}{\bibfnamefont{H.-S.} \bibnamefont{Lee}},
  \bibinfo{journal}{JHEP} \textbf{\bibinfo{volume}{02}}, \bibinfo{pages}{031}
  (\bibinfo{year}{2017}), \eprint{1606.09317}.

\bibitem[{\citenamefont{Chang et~al.}(2017)\citenamefont{Chang, Essig, and
  McDermott}}]{Chang:2016ntp}
\bibinfo{author}{\bibfnamefont{J.~H.} \bibnamefont{Chang}},
  \bibinfo{author}{\bibfnamefont{R.}~\bibnamefont{Essig}}, \bibnamefont{and}
  \bibinfo{author}{\bibfnamefont{S.~D.} \bibnamefont{McDermott}},
  \bibinfo{journal}{JHEP} \textbf{\bibinfo{volume}{01}}, \bibinfo{pages}{107}
  (\bibinfo{year}{2017}), \eprint{1611.03864}.

\bibitem[{\citenamefont{Ahlgren et~al.}(2013)\citenamefont{Ahlgren, Ohlsson,
  and Zhou}}]{Ahlgren:2013wba}
\bibinfo{author}{\bibfnamefont{B.}~\bibnamefont{Ahlgren}},
  \bibinfo{author}{\bibfnamefont{T.}~\bibnamefont{Ohlsson}}, \bibnamefont{and}
  \bibinfo{author}{\bibfnamefont{S.}~\bibnamefont{Zhou}},
  \bibinfo{journal}{Phys. Rev. Lett.} \textbf{\bibinfo{volume}{111}},
  \bibinfo{pages}{199001} (\bibinfo{year}{2013}), \eprint{1309.0991}.

\bibitem[{\citenamefont{Williams et~al.}(2011)\citenamefont{Williams, Burgess,
  Maharana, and Quevedo}}]{Williams:2011qb}
\bibinfo{author}{\bibfnamefont{M.}~\bibnamefont{Williams}},
  \bibinfo{author}{\bibfnamefont{C.~P.} \bibnamefont{Burgess}},
  \bibinfo{author}{\bibfnamefont{A.}~\bibnamefont{Maharana}}, \bibnamefont{and}
  \bibinfo{author}{\bibfnamefont{F.}~\bibnamefont{Quevedo}},
  \bibinfo{journal}{JHEP} \textbf{\bibinfo{volume}{08}}, \bibinfo{pages}{106}
  (\bibinfo{year}{2011}), \eprint{1103.4556}.

\bibitem[{\citenamefont{Mayle et~al.}(1988)\citenamefont{Mayle, Wilson, Ellis,
  Olive, Schramm, and Steigman}}]{Mayle:1987as}
\bibinfo{author}{\bibfnamefont{R.}~\bibnamefont{Mayle}},
  \bibinfo{author}{\bibfnamefont{J.~R.} \bibnamefont{Wilson}},
  \bibinfo{author}{\bibfnamefont{J.~R.} \bibnamefont{Ellis}},
  \bibinfo{author}{\bibfnamefont{K.~A.} \bibnamefont{Olive}},
  \bibinfo{author}{\bibfnamefont{D.~N.} \bibnamefont{Schramm}},
  \bibnamefont{and} \bibinfo{author}{\bibfnamefont{G.}~\bibnamefont{Steigman}},
  \bibinfo{journal}{Phys. Lett.} \textbf{\bibinfo{volume}{B203}},
  \bibinfo{pages}{188} (\bibinfo{year}{1988}).

\bibitem[{\citenamefont{Raffelt and Seckel}(1988)}]{Raffelt:1987yt}
\bibinfo{author}{\bibfnamefont{G.}~\bibnamefont{Raffelt}} \bibnamefont{and}
  \bibinfo{author}{\bibfnamefont{D.}~\bibnamefont{Seckel}},
  \bibinfo{journal}{Phys. Rev. Lett.} \textbf{\bibinfo{volume}{60}},
  \bibinfo{pages}{1793} (\bibinfo{year}{1988}).

\bibitem[{\citenamefont{Turner}(1988)}]{Turner:1987by}
\bibinfo{author}{\bibfnamefont{M.~S.} \bibnamefont{Turner}},
  \bibinfo{journal}{Phys. Rev. Lett.} \textbf{\bibinfo{volume}{60}},
  \bibinfo{pages}{1797} (\bibinfo{year}{1988}).

\bibitem[{\citenamefont{Bjorken et~al.}(2009)\citenamefont{Bjorken, Essig,
  Schuster, and Toro}}]{Bjorken:2009mm}
\bibinfo{author}{\bibfnamefont{J.~D.} \bibnamefont{Bjorken}},
  \bibinfo{author}{\bibfnamefont{R.}~\bibnamefont{Essig}},
  \bibinfo{author}{\bibfnamefont{P.}~\bibnamefont{Schuster}}, \bibnamefont{and}
  \bibinfo{author}{\bibfnamefont{N.}~\bibnamefont{Toro}},
  \bibinfo{journal}{Phys. Rev.} \textbf{\bibinfo{volume}{D80}},
  \bibinfo{pages}{075018} (\bibinfo{year}{2009}), \eprint{0906.0580}.

\bibitem[{\citenamefont{Vilain et~al.}(1993)}]{Vilain:1993kd}
\bibinfo{author}{\bibfnamefont{P.}~\bibnamefont{Vilain}} \bibnamefont{et~al.}
  (\bibinfo{collaboration}{CHARM-II}), \bibinfo{journal}{Phys. Lett.}
  \textbf{\bibinfo{volume}{B302}}, \bibinfo{pages}{351} (\bibinfo{year}{1993}).

\bibitem[{\citenamefont{Bjorken et~al.}(1988)\citenamefont{Bjorken, Ecklund,
  Nelson, Abashian, Church, Lu, Mo, Nunamaker, and Rassmann}}]{Bjorken:1988as}
\bibinfo{author}{\bibfnamefont{J.~D.} \bibnamefont{Bjorken}},
  \bibinfo{author}{\bibfnamefont{S.}~\bibnamefont{Ecklund}},
  \bibinfo{author}{\bibfnamefont{W.~R.} \bibnamefont{Nelson}},
  \bibinfo{author}{\bibfnamefont{A.}~\bibnamefont{Abashian}},
  \bibinfo{author}{\bibfnamefont{C.}~\bibnamefont{Church}},
  \bibinfo{author}{\bibfnamefont{B.}~\bibnamefont{Lu}},
  \bibinfo{author}{\bibfnamefont{L.~W.} \bibnamefont{Mo}},
  \bibinfo{author}{\bibfnamefont{T.~A.} \bibnamefont{Nunamaker}},
  \bibnamefont{and} \bibinfo{author}{\bibfnamefont{P.}~\bibnamefont{Rassmann}},
  \bibinfo{journal}{Phys. Rev.} \textbf{\bibinfo{volume}{D38}},
  \bibinfo{pages}{3375} (\bibinfo{year}{1988}).

\bibitem[{\citenamefont{Riordan et~al.}(1987)}]{Riordan:1987aw}
\bibinfo{author}{\bibfnamefont{E.~M.} \bibnamefont{Riordan}}
  \bibnamefont{et~al.}, \bibinfo{journal}{Phys. Rev. Lett.}
  \textbf{\bibinfo{volume}{59}}, \bibinfo{pages}{755} (\bibinfo{year}{1987}).

\bibitem[{\citenamefont{Bross et~al.}(1991)\citenamefont{Bross, Crisler,
  Pordes, Volk, Errede, and Wrbanek}}]{Bross:1989mp}
\bibinfo{author}{\bibfnamefont{A.}~\bibnamefont{Bross}},
  \bibinfo{author}{\bibfnamefont{M.}~\bibnamefont{Crisler}},
  \bibinfo{author}{\bibfnamefont{S.~H.} \bibnamefont{Pordes}},
  \bibinfo{author}{\bibfnamefont{J.}~\bibnamefont{Volk}},
  \bibinfo{author}{\bibfnamefont{S.}~\bibnamefont{Errede}}, \bibnamefont{and}
  \bibinfo{author}{\bibfnamefont{J.}~\bibnamefont{Wrbanek}},
  \bibinfo{journal}{Phys. Rev. Lett.} \textbf{\bibinfo{volume}{67}},
  \bibinfo{pages}{2942} (\bibinfo{year}{1991}).

\bibitem[{\citenamefont{Fradette et~al.}(2014)\citenamefont{Fradette, Pospelov,
  Pradler, and Ritz}}]{Fradette:2014sza}
\bibinfo{author}{\bibfnamefont{A.}~\bibnamefont{Fradette}},
  \bibinfo{author}{\bibfnamefont{M.}~\bibnamefont{Pospelov}},
  \bibinfo{author}{\bibfnamefont{J.}~\bibnamefont{Pradler}}, \bibnamefont{and}
  \bibinfo{author}{\bibfnamefont{A.}~\bibnamefont{Ritz}},
  \bibinfo{journal}{Phys. Rev.} \textbf{\bibinfo{volume}{D90}},
  \bibinfo{pages}{035022} (\bibinfo{year}{2014}), \eprint{1407.0993}.

\bibitem[{\citenamefont{Berger et~al.}(2016)\citenamefont{Berger, Jedamzik, and
  Walker}}]{Berger:2016vxi}
\bibinfo{author}{\bibfnamefont{J.}~\bibnamefont{Berger}},
  \bibinfo{author}{\bibfnamefont{K.}~\bibnamefont{Jedamzik}}, \bibnamefont{and}
  \bibinfo{author}{\bibfnamefont{D.~G.~E.} \bibnamefont{Walker}},
  \bibinfo{journal}{JCAP} \textbf{\bibinfo{volume}{1611}}, \bibinfo{pages}{032}
  (\bibinfo{year}{2016}), \eprint{1605.07195}.

\bibitem[{\citenamefont{Huang et~al.}(2018)\citenamefont{Huang, Ohlsson, and
  Zhou}}]{Huang:2017egl}
\bibinfo{author}{\bibfnamefont{G.-y.} \bibnamefont{Huang}},
  \bibinfo{author}{\bibfnamefont{T.}~\bibnamefont{Ohlsson}}, \bibnamefont{and}
  \bibinfo{author}{\bibfnamefont{S.}~\bibnamefont{Zhou}},
  \bibinfo{journal}{Phys. Rev.} \textbf{\bibinfo{volume}{D97}},
  \bibinfo{pages}{075009} (\bibinfo{year}{2018}), \eprint{1712.04792}.

\bibitem[{\citenamefont{Heeck and Teresi}(2016)}]{Heeck:2016oda}
\bibinfo{author}{\bibfnamefont{J.}~\bibnamefont{Heeck}} \bibnamefont{and}
  \bibinfo{author}{\bibfnamefont{D.}~\bibnamefont{Teresi}},
  \bibinfo{journal}{Phys. Rev.} \textbf{\bibinfo{volume}{D94}},
  \bibinfo{pages}{095024} (\bibinfo{year}{2016}), \eprint{1609.03594}.

\bibitem[{\citenamefont{Weldon}(1982)}]{Weldon:1982bn}
\bibinfo{author}{\bibfnamefont{H.~A.} \bibnamefont{Weldon}},
  \bibinfo{journal}{Phys. Rev.} \textbf{\bibinfo{volume}{D26}},
  \bibinfo{pages}{2789} (\bibinfo{year}{1982}).

\bibitem[{\citenamefont{Sigl and Raffelt}(1993)}]{Sigl:1992fn}
\bibinfo{author}{\bibfnamefont{G.}~\bibnamefont{Sigl}} \bibnamefont{and}
  \bibinfo{author}{\bibfnamefont{G.}~\bibnamefont{Raffelt}},
  \bibinfo{journal}{Nucl. Phys.} \textbf{\bibinfo{volume}{B406}},
  \bibinfo{pages}{423} (\bibinfo{year}{1993}).

\bibitem[{\citenamefont{McDonald}(2002)}]{McDonald:2001vt}
\bibinfo{author}{\bibfnamefont{J.}~\bibnamefont{McDonald}},
  \bibinfo{journal}{Phys. Rev. Lett.} \textbf{\bibinfo{volume}{88}},
  \bibinfo{pages}{091304} (\bibinfo{year}{2002}), \eprint{hep-ph/0106249}.

\bibitem[{\citenamefont{Hall et~al.}(2010)\citenamefont{Hall, Jedamzik,
  March-Russell, and West}}]{Hall:2009bx}
\bibinfo{author}{\bibfnamefont{L.~J.} \bibnamefont{Hall}},
  \bibinfo{author}{\bibfnamefont{K.}~\bibnamefont{Jedamzik}},
  \bibinfo{author}{\bibfnamefont{J.}~\bibnamefont{March-Russell}},
  \bibnamefont{and} \bibinfo{author}{\bibfnamefont{S.~M.} \bibnamefont{West}},
  \bibinfo{journal}{JHEP} \textbf{\bibinfo{volume}{03}}, \bibinfo{pages}{080}
  (\bibinfo{year}{2010}), \eprint{0911.1120}.

\bibitem[{\citenamefont{Dine and Fischler}(1983)}]{Dine:1982ah}
\bibinfo{author}{\bibfnamefont{M.}~\bibnamefont{Dine}} \bibnamefont{and}
  \bibinfo{author}{\bibfnamefont{W.}~\bibnamefont{Fischler}},
  \bibinfo{journal}{Phys. Lett.} \textbf{\bibinfo{volume}{B120}},
  \bibinfo{pages}{137} (\bibinfo{year}{1983}).

\bibitem[{\citenamefont{Branco et~al.}(2012)\citenamefont{Branco, Ferreira,
  Lavoura, Rebelo, Sher, and Silva}}]{Branco:2011iw}
\bibinfo{author}{\bibfnamefont{G.~C.} \bibnamefont{Branco}},
  \bibinfo{author}{\bibfnamefont{P.~M.} \bibnamefont{Ferreira}},
  \bibinfo{author}{\bibfnamefont{L.}~\bibnamefont{Lavoura}},
  \bibinfo{author}{\bibfnamefont{M.~N.} \bibnamefont{Rebelo}},
  \bibinfo{author}{\bibfnamefont{M.}~\bibnamefont{Sher}}, \bibnamefont{and}
  \bibinfo{author}{\bibfnamefont{J.~P.} \bibnamefont{Silva}},
  \bibinfo{journal}{Phys. Rept.} \textbf{\bibinfo{volume}{516}},
  \bibinfo{pages}{1} (\bibinfo{year}{2012}), \eprint{1106.0034}.

\bibitem[{\citenamefont{Espriu et~al.}(2015)\citenamefont{Espriu, Mescia, and
  Renau}}]{Espriu:2015mfa}
\bibinfo{author}{\bibfnamefont{D.}~\bibnamefont{Espriu}},
  \bibinfo{author}{\bibfnamefont{F.}~\bibnamefont{Mescia}}, \bibnamefont{and}
  \bibinfo{author}{\bibfnamefont{A.}~\bibnamefont{Renau}},
  \bibinfo{journal}{Phys. Rev.} \textbf{\bibinfo{volume}{D92}},
  \bibinfo{pages}{095013} (\bibinfo{year}{2015}), \eprint{1503.02953}.

\bibitem[{\citenamefont{Raffelt}(1999)}]{Raffelt:1999tx}
\bibinfo{author}{\bibfnamefont{G.~G.} \bibnamefont{Raffelt}},
  \bibinfo{journal}{Ann. Rev. Nucl. Part. Sci.} \textbf{\bibinfo{volume}{49}},
  \bibinfo{pages}{163} (\bibinfo{year}{1999}), \eprint{hep-ph/9903472}.

\bibitem[{\citenamefont{Preskill et~al.}(1983)\citenamefont{Preskill, Wise, and
  Wilczek}}]{Preskill:1982cy}
\bibinfo{author}{\bibfnamefont{J.}~\bibnamefont{Preskill}},
  \bibinfo{author}{\bibfnamefont{M.~B.} \bibnamefont{Wise}}, \bibnamefont{and}
  \bibinfo{author}{\bibfnamefont{F.}~\bibnamefont{Wilczek}},
  \bibinfo{journal}{Phys. Lett.} \textbf{\bibinfo{volume}{B120}},
  \bibinfo{pages}{127} (\bibinfo{year}{1983}).

\bibitem[{\citenamefont{Abbott and Sikivie}(1983)}]{Abbott:1982af}
\bibinfo{author}{\bibfnamefont{L.~F.} \bibnamefont{Abbott}} \bibnamefont{and}
  \bibinfo{author}{\bibfnamefont{P.}~\bibnamefont{Sikivie}},
  \bibinfo{journal}{Phys. Lett.} \textbf{\bibinfo{volume}{B120}},
  \bibinfo{pages}{133} (\bibinfo{year}{1983}).

\bibitem[{\citenamefont{Kawasaki and Nakayama}(2013)}]{Kawasaki:2013ae}
\bibinfo{author}{\bibfnamefont{M.}~\bibnamefont{Kawasaki}} \bibnamefont{and}
  \bibinfo{author}{\bibfnamefont{K.}~\bibnamefont{Nakayama}},
  \bibinfo{journal}{Ann. Rev. Nucl. Part. Sci.} \textbf{\bibinfo{volume}{63}},
  \bibinfo{pages}{69} (\bibinfo{year}{2013}), \eprint{1301.1123}.

\bibitem[{\citenamefont{Marsh}(2016)}]{Marsh:2015xka}
\bibinfo{author}{\bibfnamefont{D.~J.~E.} \bibnamefont{Marsh}},
  \bibinfo{journal}{Phys. Rept.} \textbf{\bibinfo{volume}{643}},
  \bibinfo{pages}{1} (\bibinfo{year}{2016}), \eprint{1510.07633}.

\bibitem[{\citenamefont{Weinberg}(2013)}]{Weinberg:2013kea}
\bibinfo{author}{\bibfnamefont{S.}~\bibnamefont{Weinberg}},
  \bibinfo{journal}{Phys. Rev. Lett.} \textbf{\bibinfo{volume}{110}},
  \bibinfo{pages}{241301} (\bibinfo{year}{2013}), \eprint{1305.1971}.

\bibitem[{\citenamefont{Graf and Steffen}(2011)}]{Graf:2010tv}
\bibinfo{author}{\bibfnamefont{P.}~\bibnamefont{Graf}} \bibnamefont{and}
  \bibinfo{author}{\bibfnamefont{F.~D.} \bibnamefont{Steffen}},
  \bibinfo{journal}{Phys. Rev.} \textbf{\bibinfo{volume}{D83}},
  \bibinfo{pages}{075011} (\bibinfo{year}{2011}), \eprint{1008.4528}.

\bibitem[{\citenamefont{Salvio et~al.}(2014)\citenamefont{Salvio, Strumia, and
  Xue}}]{Salvio:2013iaa}
\bibinfo{author}{\bibfnamefont{A.}~\bibnamefont{Salvio}},
  \bibinfo{author}{\bibfnamefont{A.}~\bibnamefont{Strumia}}, \bibnamefont{and}
  \bibinfo{author}{\bibfnamefont{W.}~\bibnamefont{Xue}},
  \bibinfo{journal}{JCAP} \textbf{\bibinfo{volume}{1401}}, \bibinfo{pages}{011}
  (\bibinfo{year}{2014}), \eprint{1310.6982}.

\bibitem[{\citenamefont{Clarke and Volkas}(2016)}]{Clarke:2015bea}
\bibinfo{author}{\bibfnamefont{J.~D.} \bibnamefont{Clarke}} \bibnamefont{and}
  \bibinfo{author}{\bibfnamefont{R.~R.} \bibnamefont{Volkas}},
  \bibinfo{journal}{Phys. Rev.} \textbf{\bibinfo{volume}{D93}},
  \bibinfo{pages}{035001} (\bibinfo{year}{2016}), \bibinfo{note}{[Phys.
  Rev.D93,035001(2016)]}, \eprint{1509.07243}.

\bibitem[{\citenamefont{Kim}(1979)}]{Kim:1979if}
\bibinfo{author}{\bibfnamefont{J.~E.} \bibnamefont{Kim}},
  \bibinfo{journal}{Phys. Rev. Lett.} \textbf{\bibinfo{volume}{43}},
  \bibinfo{pages}{103} (\bibinfo{year}{1979}).

\bibitem[{\citenamefont{Shifman et~al.}(1980)\citenamefont{Shifman, Vainshtein,
  and Zakharov}}]{Shifman:1979if}
\bibinfo{author}{\bibfnamefont{M.~A.} \bibnamefont{Shifman}},
  \bibinfo{author}{\bibfnamefont{A.~I.} \bibnamefont{Vainshtein}},
  \bibnamefont{and} \bibinfo{author}{\bibfnamefont{V.~I.}
  \bibnamefont{Zakharov}}, \bibinfo{journal}{Nucl. Phys.}
  \textbf{\bibinfo{volume}{B166}}, \bibinfo{pages}{493} (\bibinfo{year}{1980}).

\bibitem[{\citenamefont{Ballesteros
  et~al.}(2017{\natexlab{a}})\citenamefont{Ballesteros, Redondo, Ringwald, and
  Tamarit}}]{Ballesteros:2016euj}
\bibinfo{author}{\bibfnamefont{G.}~\bibnamefont{Ballesteros}},
  \bibinfo{author}{\bibfnamefont{J.}~\bibnamefont{Redondo}},
  \bibinfo{author}{\bibfnamefont{A.}~\bibnamefont{Ringwald}}, \bibnamefont{and}
  \bibinfo{author}{\bibfnamefont{C.}~\bibnamefont{Tamarit}},
  \bibinfo{journal}{Phys. Rev. Lett.} \textbf{\bibinfo{volume}{118}},
  \bibinfo{pages}{071802} (\bibinfo{year}{2017}{\natexlab{a}}),
  \eprint{1608.05414}.

\bibitem[{\citenamefont{Ballesteros
  et~al.}(2017{\natexlab{b}})\citenamefont{Ballesteros, Redondo, Ringwald, and
  Tamarit}}]{Ballesteros:2016xej}
\bibinfo{author}{\bibfnamefont{G.}~\bibnamefont{Ballesteros}},
  \bibinfo{author}{\bibfnamefont{J.}~\bibnamefont{Redondo}},
  \bibinfo{author}{\bibfnamefont{A.}~\bibnamefont{Ringwald}}, \bibnamefont{and}
  \bibinfo{author}{\bibfnamefont{C.}~\bibnamefont{Tamarit}},
  \bibinfo{journal}{JCAP} \textbf{\bibinfo{volume}{1708}}, \bibinfo{pages}{001}
  (\bibinfo{year}{2017}{\natexlab{b}}), \eprint{1610.01639}.

\bibitem[{\citenamefont{Asaka et~al.}(2006)\citenamefont{Asaka, Laine, and
  Shaposhnikov}}]{Asaka:2006rw}
\bibinfo{author}{\bibfnamefont{T.}~\bibnamefont{Asaka}},
  \bibinfo{author}{\bibfnamefont{M.}~\bibnamefont{Laine}}, \bibnamefont{and}
  \bibinfo{author}{\bibfnamefont{M.}~\bibnamefont{Shaposhnikov}},
  \bibinfo{journal}{JHEP} \textbf{\bibinfo{volume}{06}}, \bibinfo{pages}{053}
  (\bibinfo{year}{2006}), \eprint{hep-ph/0605209}.

\bibitem[{\citenamefont{Cline et~al.}(1993)\citenamefont{Cline, Kainulainen,
  and Olive}}]{Cline:1993ht}
\bibinfo{author}{\bibfnamefont{J.~M.} \bibnamefont{Cline}},
  \bibinfo{author}{\bibfnamefont{K.}~\bibnamefont{Kainulainen}},
  \bibnamefont{and} \bibinfo{author}{\bibfnamefont{K.~A.} \bibnamefont{Olive}},
  \bibinfo{journal}{Astropart. Phys.} \textbf{\bibinfo{volume}{1}},
  \bibinfo{pages}{387} (\bibinfo{year}{1993}), \eprint{hep-ph/9304229}.

\bibitem[{\citenamefont{Gu and Sarkar}(2011)}]{Gu:2009hn}
\bibinfo{author}{\bibfnamefont{P.-H.} \bibnamefont{Gu}} \bibnamefont{and}
  \bibinfo{author}{\bibfnamefont{U.}~\bibnamefont{Sarkar}},
  \bibinfo{journal}{Eur. Phys. J.} \textbf{\bibinfo{volume}{C71}},
  \bibinfo{pages}{1560} (\bibinfo{year}{2011}), \eprint{0909.5468}.

\bibitem[{\citenamefont{Dev et~al.}(2018)\citenamefont{Dev, Mohapatra, and
  Zhang}}]{Dev:2017xry}
\bibinfo{author}{\bibfnamefont{P.~S.~B.} \bibnamefont{Dev}},
  \bibinfo{author}{\bibfnamefont{R.~N.} \bibnamefont{Mohapatra}},
  \bibnamefont{and} \bibinfo{author}{\bibfnamefont{Y.}~\bibnamefont{Zhang}},
  \bibinfo{journal}{JHEP} \textbf{\bibinfo{volume}{03}}, \bibinfo{pages}{122}
  (\bibinfo{year}{2018}), \eprint{1711.07634}.

\bibitem[{\citenamefont{Akhmedov et~al.}(1993)\citenamefont{Akhmedov,
  Berezhiani, Mohapatra, and Senjanovic}}]{Akhmedov:1992hi}
\bibinfo{author}{\bibfnamefont{E.~K.} \bibnamefont{Akhmedov}},
  \bibinfo{author}{\bibfnamefont{Z.~G.} \bibnamefont{Berezhiani}},
  \bibinfo{author}{\bibfnamefont{R.~N.} \bibnamefont{Mohapatra}},
  \bibnamefont{and}
  \bibinfo{author}{\bibfnamefont{G.}~\bibnamefont{Senjanovic}},
  \bibinfo{journal}{Phys. Lett.} \textbf{\bibinfo{volume}{B299}},
  \bibinfo{pages}{90} (\bibinfo{year}{1993}), \eprint{hep-ph/9209285}.

\bibitem[{\citenamefont{Rothstein et~al.}(1993)\citenamefont{Rothstein, Babu,
  and Seckel}}]{Rothstein:1992rh}
\bibinfo{author}{\bibfnamefont{I.~Z.} \bibnamefont{Rothstein}},
  \bibinfo{author}{\bibfnamefont{K.~S.} \bibnamefont{Babu}}, \bibnamefont{and}
  \bibinfo{author}{\bibfnamefont{D.}~\bibnamefont{Seckel}},
  \bibinfo{journal}{Nucl. Phys.} \textbf{\bibinfo{volume}{B403}},
  \bibinfo{pages}{725} (\bibinfo{year}{1993}), \eprint{hep-ph/9301213}.

\bibitem[{\citenamefont{Hebecker et~al.}(2017)\citenamefont{Hebecker, Mangat,
  Theisen, and Witkowski}}]{Hebecker:2016dsw}
\bibinfo{author}{\bibfnamefont{A.}~\bibnamefont{Hebecker}},
  \bibinfo{author}{\bibfnamefont{P.}~\bibnamefont{Mangat}},
  \bibinfo{author}{\bibfnamefont{S.}~\bibnamefont{Theisen}}, \bibnamefont{and}
  \bibinfo{author}{\bibfnamefont{L.~T.} \bibnamefont{Witkowski}},
  \bibinfo{journal}{JHEP} \textbf{\bibinfo{volume}{02}}, \bibinfo{pages}{097}
  (\bibinfo{year}{2017}), \eprint{1607.06814}.

\bibitem[{\citenamefont{Alonso and Urbano}(2017)}]{Alonso:2017avz}
\bibinfo{author}{\bibfnamefont{R.}~\bibnamefont{Alonso}} \bibnamefont{and}
  \bibinfo{author}{\bibfnamefont{A.}~\bibnamefont{Urbano}}
  (\bibinfo{year}{2017}), \eprint{1706.07415}.

\bibitem[{\citenamefont{Lee}(1988)}]{Lee:1988ge}
\bibinfo{author}{\bibfnamefont{K.-M.} \bibnamefont{Lee}},
  \bibinfo{journal}{Phys. Rev. Lett.} \textbf{\bibinfo{volume}{61}},
  \bibinfo{pages}{263} (\bibinfo{year}{1988}).

\bibitem[{\citenamefont{Hill and Ross}(1988)}]{Hill:1988bu}
\bibinfo{author}{\bibfnamefont{C.~T.} \bibnamefont{Hill}} \bibnamefont{and}
  \bibinfo{author}{\bibfnamefont{G.~G.} \bibnamefont{Ross}},
  \bibinfo{journal}{Nucl. Phys.} \textbf{\bibinfo{volume}{B311}},
  \bibinfo{pages}{253} (\bibinfo{year}{1988}).

\bibitem[{\citenamefont{Frigerio et~al.}(2011)\citenamefont{Frigerio, Hambye,
  and Masso}}]{Frigerio:2011in}
\bibinfo{author}{\bibfnamefont{M.}~\bibnamefont{Frigerio}},
  \bibinfo{author}{\bibfnamefont{T.}~\bibnamefont{Hambye}}, \bibnamefont{and}
  \bibinfo{author}{\bibfnamefont{E.}~\bibnamefont{Masso}},
  \bibinfo{journal}{Phys. Rev.} \textbf{\bibinfo{volume}{X1}},
  \bibinfo{pages}{021026} (\bibinfo{year}{2011}), \eprint{1107.4564}.

\bibitem[{\citenamefont{Lattanzi and Valle}(2007)}]{Lattanzi:2007ux}
\bibinfo{author}{\bibfnamefont{M.}~\bibnamefont{Lattanzi}} \bibnamefont{and}
  \bibinfo{author}{\bibfnamefont{J.~W.~F.} \bibnamefont{Valle}},
  \bibinfo{journal}{Phys. Rev. Lett.} \textbf{\bibinfo{volume}{99}},
  \bibinfo{pages}{121301} (\bibinfo{year}{2007}), \eprint{0705.2406}.

\bibitem[{\citenamefont{Lattanzi et~al.}(2014)\citenamefont{Lattanzi, Lineros,
  and Taoso}}]{Lattanzi:2014mia}
\bibinfo{author}{\bibfnamefont{M.}~\bibnamefont{Lattanzi}},
  \bibinfo{author}{\bibfnamefont{R.~A.} \bibnamefont{Lineros}},
  \bibnamefont{and} \bibinfo{author}{\bibfnamefont{M.}~\bibnamefont{Taoso}},
  \bibinfo{journal}{New J. Phys.} \textbf{\bibinfo{volume}{16}},
  \bibinfo{pages}{125012} (\bibinfo{year}{2014}), \eprint{1406.0004}.

\bibitem[{\citenamefont{Choi and Santamaria}(1990)}]{Choi:1989hi}
\bibinfo{author}{\bibfnamefont{K.}~\bibnamefont{Choi}} \bibnamefont{and}
  \bibinfo{author}{\bibfnamefont{A.}~\bibnamefont{Santamaria}},
  \bibinfo{journal}{Phys. Rev.} \textbf{\bibinfo{volume}{D42}},
  \bibinfo{pages}{293} (\bibinfo{year}{1990}).

\bibitem[{\citenamefont{Kusenko}(2006)}]{Kusenko:2006rh}
\bibinfo{author}{\bibfnamefont{A.}~\bibnamefont{Kusenko}},
  \bibinfo{journal}{Phys. Rev. Lett.} \textbf{\bibinfo{volume}{97}},
  \bibinfo{pages}{241301} (\bibinfo{year}{2006}), \eprint{hep-ph/0609081}.

\bibitem[{\citenamefont{Petraki and Kusenko}(2008)}]{Petraki:2007gq}
\bibinfo{author}{\bibfnamefont{K.}~\bibnamefont{Petraki}} \bibnamefont{and}
  \bibinfo{author}{\bibfnamefont{A.}~\bibnamefont{Kusenko}},
  \bibinfo{journal}{Phys. Rev.} \textbf{\bibinfo{volume}{D77}},
  \bibinfo{pages}{065014} (\bibinfo{year}{2008}), \eprint{0711.4646}.

\bibitem[{\citenamefont{Sekiya et~al.}(2015)\citenamefont{Sekiya, Yamasaki, and
  Mitsuda}}]{Sekiya:2015jsa}
\bibinfo{author}{\bibfnamefont{N.}~\bibnamefont{Sekiya}},
  \bibinfo{author}{\bibfnamefont{N.~Y.} \bibnamefont{Yamasaki}},
  \bibnamefont{and} \bibinfo{author}{\bibfnamefont{K.}~\bibnamefont{Mitsuda}},
  \bibinfo{journal}{Publ. Astron. Soc. Jap.}  (\bibinfo{year}{2015}),
  \eprint{1504.02826}.

\bibitem[{\citenamefont{Boyarsky et~al.}(2008)\citenamefont{Boyarsky, Malyshev,
  Neronov, and Ruchayskiy}}]{Boyarsky:2007ge}
\bibinfo{author}{\bibfnamefont{A.}~\bibnamefont{Boyarsky}},
  \bibinfo{author}{\bibfnamefont{D.}~\bibnamefont{Malyshev}},
  \bibinfo{author}{\bibfnamefont{A.}~\bibnamefont{Neronov}}, \bibnamefont{and}
  \bibinfo{author}{\bibfnamefont{O.}~\bibnamefont{Ruchayskiy}},
  \bibinfo{journal}{Mon. Not. Roy. Astron. Soc.}
  \textbf{\bibinfo{volume}{387}}, \bibinfo{pages}{1345} (\bibinfo{year}{2008}),
  \eprint{0710.4922}.

\bibitem[{\citenamefont{Baumholzer et~al.}(2018)\citenamefont{Baumholzer,
  Brdar, and Schwaller}}]{Baumholzer:2018sfb}
\bibinfo{author}{\bibfnamefont{S.}~\bibnamefont{Baumholzer}},
  \bibinfo{author}{\bibfnamefont{V.}~\bibnamefont{Brdar}}, \bibnamefont{and}
  \bibinfo{author}{\bibfnamefont{P.}~\bibnamefont{Schwaller}}
  (\bibinfo{year}{2018}), \eprint{1806.06864}.

\end{thebibliography}

\end{document}